# CHARACTERISING THE DIFFERENCE BETWEEN COMPLEX ADAPTIVE AND CONVENTIONAL COMBAT MODELS

M K Lauren

November 1999

Abstract:

Here an attempt is made to quantitatively demonstrate the difference between the statistics of a complex adaptive combat model and those produced by conventional combat models. The work will show that complex adaptive models of combat may give rise to "turbulent" dynamics, which emerge once the battlefield is no longer "linear", i.e. once military formations no longer form ordered lines or columns. This is done using a cellular automaton model. This model exhibits a high degree of complexity, which leads to a rich variety of behaviour. Conventional statistical methods fail to adequately explain or capture this richness. In this report, particular attention is paid to the properties of the attrition function. It is found that this function is discontinuous and possesses non-linear properties such as the clustering of casualties, scaling of the statistical moments of the data and fat-tailed probability distributions. Fractal methods are found to be capable of quantifying these properties. A particularly significant result is the non-linearity of the pay-off of improved weapons performance. This contrasts with conventional models, where the attrition rate usually depends in a near-linear way on kill probability. Additionally, the model is found to possess at least two attractive states, each with their own set of statistical properties. A transition exists between the two cases in some situations. This transition appears to behave in an analogous way to the transition between laminar and turbulent states in fluid dynamics, i.e. the transition point is unpredictable and rapid. Here, the "gradients" which develop due to imbalances in the strengths of the opposing forces play the central role in determining how the combat evolves (due to its adaptive nature), while the degree to which each force sticks together plays an analogous role to viscosity.

Defence Operational Support Technology Establishment
Auckland, New Zealand

## EXECUTIVE SUMMARY

### Background

The New Zealand Army is keeping a close eye on trends in land warfare as part of the Tekapo Manoeuvres modernisation programme. It is hoped that operations analysis may be used to determine how the Army should structure itself to cope in present and future regions of conflict.

Current operations analysis methods and wargaming tools typically place heavy emphasis on weapons capabilities, but often ignore (or at least, only model in terms of "fudge factors") a wealth of intangibles that are known to be of importance on the battlefield, such as troop quality, training and morale. Consequently, improving the capability of a weapon system within such a model tends to improve its performance in a linear way. For example, doubling the kill probability of a weapon will roughly double the attrition rate of its target. However, if the targets adapt their tactics to counter the increased threat of the weapon, the pay-off of this improvement may be much lower than otherwise expected. Such an effect is "non-linear" mathematically speaking, since in such a circumstance one may receive a much smaller pay off from the improvement in kill probability than would "intuitively" be expected given the level of the weapon improvement. The ability of one side to adapt to counter the capabilities of the other is often referred to as a "co-evolving landscape" of tactical behaviour.

The Defence Operational Technology Support Establishment (DOTSE) along with the Capability, Analysis and Doctrine (CAD) and Development (Dev) Branches of the NZ Army have recognised that much more can be gained from the technology currently available if a good understanding exists of the non-linearities which arise with such co-evolving tactics. In particular, it is hoped that these non-linearities may be exploited to gain a much bigger pay-off than might be expected given the costs and level of improvements to the weapons. This may result from using the right equipment in the right combinations with the right kind of tactics to cause a much greater impact on the opponent than any of these pieces of equipment would in isolation.

To explore these possibilities, we require models which treat combat as a non-linear, co-evolving system. Though this was recognised by DOTSE earlier this year, it was not until a programme of collaborative research began with the United States Marine Corp Combat Development Command (USMC CDC) that such a model became available here. The model, ISAAC, is a "cellular automaton" model. Such models possess the ability to adapt to local situations without the explicit instruction of the modeller.

Although relatively simple, ISAAC has the ability to produce a rich variety of behaviour. And though there is potential to improve the realism of the model, at this point in time it may be more useful to attempt to understand its properties in its current form. This may give a clearer picture of the kinds of behaviour the non-linearities in the model lead to, without obscuring this with complicated details of weapons systems. That is, ISAAC is suited to exploring concepts rather than specific equipment types.

The results presented are intended to reveal something of the behaviour of models such as ISAAC. In particular, it is hoped that not only will something be learned about the statistics of such models, but that we may gain insights into how they should be used.

While it is clear to a casual observer that the ISAAC model is capable of exhibiting a rich variety of behaviour, it is not so clear how this behaviour can be characterised. This allows sceptics to question whether the ISAAC methodology really produces useful insights which are significantly different from those which may be gained from conventional combat models. This report demonstrates how such differences can be characterised, and shows that the "personality" rules within ISAAC indeed have a significant effect, particularly on the estimation of risk.

Though this report contains some mathematical concepts, we encourage those not familiar with the relevant statistical theory to at least read the "descriptive" text, which expresses these concepts in a non-mathematical way.

The report is intended as an overview of some of the mathematical peculiarities models like ISAAC present. Unfortunately, in order that this document not become too long and unwieldy, certain aspects have not been developed in sufficient detail to satisfy all readers. As such, it is a platform from which more detailed, succinct and specialised documents may be produced. The hope is that this further work may lead to the production of research papers that make a significant contribution to the field.

**Sponsor**

New Zealand Army.

**Aim**

To improve the understanding of, and capability to describe, the behaviour of complex adaptive combat models.

**Results**

Given two roughly equal forces, each of which is assigned the task of attacking the other, the ISAAC model was found to fight battles in one of two ways: forming lines, or using fluid, turbulent formations. The former case produces statistics which are similar to, but not exactly the same as, those of conventional combat models. The latter produces statistics which are more exotic, and certainly non-Gaussian. These "exotic" statistics do not exist in conventional models. It is also possible that the "linear" battlefield that characterises the former case can transform into the latter case during the battle in an analogous way to the transition from laminar to turbulent flow in fluid dynamics.

The different between the Gaussian-like and exotic statistics can be characterised using fractal statistics. In particular, on a turbulent, non-linear battlefield the statistics are:

• Markedly more intermittent. That is, casualties tend to come in bursts, even without explicit programming of tactics.

• The attrition rate depends on the cube root of the kill probability. This contrasts with conventional models, where doubling the kill probability of a weapon would typically double the attrition rate. This is because generally conventional models compare weapons performances keeping everything else within the scenario the same. Therefore, the "opponent" does not have the ability to react to the improvement. On the other hand, the ISAAC rules cause each side to "adjust" to the kill probability of the opponent's weapons. This is remarkable when it is remembered that there is no explicit programming of tactics in ISAAC.

• On the turbulent battlefield, the statistics possess a much greater chance of extreme events occurring than is the case for conventional models. Of particular interest is the different nature of risk for this case. For conventional models, if a situation produces on average 15 per cent casualties, say, then the statistics are such that the most common outcome from a given run is about 15 per cent casualties, with the chances of a higher or lower level falling off rapidly the further the level moves from 15 per cent. By comparison, ISAAC statistics suggest that the most common outcome of a run is 0 per cent casualties, but sometimes the casualties greatly exceed 15 per cent (i.e. "disasters" occur). Intuitively, this seems true to life, particularly when considering relatively low-risk operations such as peace keeping. In such situations, the force carrying out the operation is almost always safe, but occasionally conditions may conspire to produce extremely dangerous situations. The work here suggests ISAAC-type models may be the most suitable for modelling these kinds of events.

# Contents

# 1 Non-linearity in warfare

Advances in electronics and computer technology have brought many enhanced capabilities to the modern battlefield. Extremely long-range firepower (e.g. accurate long-range missiles, air strikes and modern artillery) combined with a high-level of situational awareness (thanks to GPS, satellite technology, and enhanced sensors), may allow war to be fought and won without relying on overwhelming numerical superiority. Consequently, traditional attrition-based models of combat described by Lanchester equations are becoming less relevant as a tool for analysing or predicting likely combat outcomes.

Emphasis must instead be placed on analysing how manoeuvre affects combat. As increasingly lethal long-range weapon systems appear on the battlefield, it becomes less attractive to mass forces for a concentrated attack, because the advantages of such a mass offensive become outweighed by the risks of presenting such a densely packed target to the opponent's long-range weapons systems. It may be argued that the trend in land warfare may be toward many small mobile units sparsely dispersed, which use their situational awareness to concentrate long-range firepower onto attractive targets as they present themselves. Even so, such a view seems to be at odds with traditional warfighting methods which are typically still employed by most of the world's armies.

In this increasingly complicated environment, defence forces have been using combat simulations for both analysis and training for a number of years. Thanks to advances in computer technology, such combat situations may be relatively easily modelled in pain-staking detail on a PC computer. This detail usually places a strong emphasis on the specifics of weapons systems. Thus assessments are often made on the basis of who has the best weapon system. Unfortunately, this ignores the wealth of intangibles which have been recognised to play vital roles in warfare. To quote Napoleon: "Morale is to the physical as three is to one".

At the same time, the tendency to emphasise manoeuvre in modern warfare has led to a recognition that combat is "non-linear". Here, we use the term non-linear with a double meaning.

Firstly, in terms of military formations. Linear warfare is based on the use of formations such as lines and columns to conduct battles. Modern weapons have forced a more dispersed approach to army deployments, while armoured transportation provides a means by which armies may manoeuvre rapidly.

In the second sense, we mean mathematical non-linearity. i.e. the whole is more (or less) than the sum of its components. For example, there are countless examples in military history of a "superior" (either numerically or technically) force being defeated by an inferior force. In such a case, we say that the superior force would win if kill potential was the only determiner (i.e. if both sides merely lined up and slogged it out to the last man). Often in these cases the supposed inferior force had a superior ability to adapt to the situation, which, combined with less-tangible attributes such as morale, allowed it to prevail. Certain military commanders (notably Alexander the Great and Napoleon Bonaparte) have been particularly adept at exploiting the non-linear nature of warfare to defeat apparently far superior opponents, without relying on superior technology.

Initial studies at DOTSE (i.e. Lauren 1999) suggest that dynamism on the battlefield is likely to lead to "clumpiness" in attrition, that is, brief bursts of engagement resulting in casualties which punctuate longer periods of few or no casualties. By contrast, the Lanchester equations assume a continuous attrition rate, so that casualties occur constantly throughout the engagement. It is known from fields such as geophysics and finance that the existence of clumpiness in data leads to statistical behaviour which can be characterised conveniently with fractal models. Data which obeys fractal statistics tends to have greater extremities, and hence their applicability to combat data has significant implications for assessing risk. That is to say, if modern "non-linear" methods of warfare do indeed produce clumpiness in the casualty rates, this affects the statistics, and hence risk.

Perhaps a good illustration of this is the Battle of Midway. Although this was a naval battle and this paper concentrates on land-operations, it is perhaps indicative of how future land wars may be fought, with the outcome resting on a few high-value elements. Because the battle was largely determined by who was able to use their situational awareness to catch the other's carriers out, such an outcome to the battle could not be explained simply by attrition (that is, could not be modelled by assuming each side continuously causes attrition on the other until one runs out of forces). Also, it may be expected that if the battle were re-run many times over, there would be a wide variance of results. It is hypothesised here (and hopefully shown) that increasing complexity on the battlefield leads to increasing variance in the outcomes. It is thus vitally important to understand the nature of these non-linearities in modern warfighting.

For land warfare, it has long been observed that there are many intangibles on the battlefield, as observed, for example, by Clausewitz last century (Beyerchen 1992).Yet traditional combat models, while capable of modelling things like weapon performance in intricate detail, appear to have little room to model these more "qualitative" aspects. In fairness, such an approach is not necessarily poor if the question at hand is which is the better weapon system from a choice of candidates. However, whether these models can provide meaningful insights into the nature of manoeuvre warfare must be questioned.

In order to characterise the non-linearities of warfare, it is necessary to use the appropriate tools. In a previous report (Lauren 1999), the author investigated the implications of using a discontinuous, clumpy function (such as a fractal) to describe attrition. The outcome of such an assumption is that the statistics of the combat show fractal traits, including fat-tailed probability distributions and scaling of the statistical moments of the combat data. If these properties do apply to combat statistics, then they must be incorporated into statistical models of combat, such as CDA's SMICC (Hall, Wright and Young, 1997), which for a given level of force provides a probability distribution for the outcomes. In fact, the same applies to any model which seeks to provide likely distributions of outcomes from predictions based on historical data, like those suggested by workers like Dupuy (1979).

Such models are typically used to reduce the problem of determining the outcome of combat from one where the behaviour of every participant in the combat must be modelled, to one where a single result is given for the entire unit from an assumed probability distribution. While the latter method models the situation less explicitly, and may therefore seem less accurate, it can be argued that if the probability distribution is based on historical data, then such a method may give a more realistic result than trying to conceive how every participant would behave. This is because such detailed combat models usually ignore human behavioural factors, such as how an individual soldier reacts under fire. For historical data, all the vagaries of human nature go into producing the data distribution, even though we may have no idea what happened in each engagement to produce that distribution.

However, the difficulty with such an historical "black-box" approach is that there is no guarantee that the historical data represents the likely outcomes of future battles, given the change in warfighting styles and technology. Even in situations where a purely statistical model may be adequate (for example, for recreating historical battles, or for some training purposes), the question becomes, what type of statistics should be used? Normal or Poisson, as expected for uncorrelated random variables; or fractal, which exhibit spatial or temporal correlations and may be used to describe complex, non-linear systems such as the weather (e.g. Tsonis and Elsner, 1989) and the sharemarket (e.g. Mandelbrot 1997).

To put it bluntly, we need some sort of new paradigm that will allow us to better understand the complexities of warfare.

Recently, much attention has been given to the non-linearity of combat (e.g. Hunt, 1998; Hoffman and Horne, 1998), and this has been explored using automaton models (e.g. Ilachinski, 1997). In the work for this report, the behaviour of the ISAAC automaton combat model is examined for evidence that the attrition rate is discontinuous and "clumpy," and the statistics examined for evidence of scaling, as suggested in Lauren (1999).

Additionally, evidence is sought that these models demonstrate behaviour which is not strongly dependent on the initial conditions, i.e. the model evolves towards certain states which are in general extremely difficult to pick from the initial conditions. Furthermore, there may be many of these states, and it is possible that a given model run may flip from one state to another. Such states are "attractors" for the dynamics of the model, and are interesting because their existence implies that the same sorts of things tend to happen in the conflict, regards of the precise paths taken to these events. Note that this is not the case for traditional physics-based combat models, which usually re-run a single pre-conceived state multiple times.

By seeking such attractors, we appeal to the notion that there exists some underlying law of combat which transcends the actual weapon systems employed, as alluded to by military historians such as Dupuy (1987). Intuitively, there are basic rules of combat which have survived the passage of time, for example, seeking to avoid being caught in a vulnerable position, while trying to put your opponent in such a situation, and hence multiply the effectiveness of your own force. Though advancements in armaments have generally meant that force deployments have become more dispersed, this basic rule still holds true.

It is the goal of work such as that presented here to end the age-old search for such a set of laws by constructing a mathematical theory which describes combat behaviour in terms of "attractive" states to which the combat dynamics evolve (attractive in the sense that the simulation evolves towards certain qualitative behaviour regardless of the initial conditions). Though it may never be possible to write down a set of equations which describe combat completely, it may be possible that the nature of combat can still be understood by finding methods for quantifying the attractive states themselves. This report sets out to find such attractive states in the ISAAC cellular automaton combat model.

There is good reason to do this, because a better understanding of the non-linear nature of warfare may lead to much greater pay-offs than may be expected from improvements in equipment alone.

## 2    The ISAAC automaton model and simulations

The cellular automaton model used in this study was the ISAAC model developed by the United States' Center for Naval Analyses. Descriptions of weapon systems within the model are simple compared with conventional combat simulation tools, and the model relies on a high level of abstraction. Yet despite this simplicity, the model is capable of demonstrating a high degree of complexity, as indeed are many such simple automaton models (e.g. Bak *et al*, 1989, 1990 ).

In this regard the model's simplicity is its strength. It allows the user to gain a clearer impression of where the complex behaviour is generated — that is, as a result of the non-linear nature of the model dynamics rather than the complexity of the model itself. Also, since entities within ISAAC are not confined to particular equipment types; the model explores concepts rather than weapons performance.

Though the terminology of the subject of complexity is still not well-established, such automaton models are generally known as "complex adaptive systems". As the name suggests, the automata react within the simulation to local circumstances. On the battlefield, it is not difficult to imagine that the individuals participating tend to react to their local circumstances, yet this fact may often be lost when those individuals are viewed as part of a unit drawn neatly on a map.

Despite their high level of physical detail, the failure of conventional combat models to describe the ability of units to react and organise themselves to fit a particular situation is a serious short-coming in representing reality. This is particularly so because these conventional models consequently fail to explore the many possible variations in the way a given battle can evolve each up (which in general is not practical, due to the very large number of possible ways to fight a given battle). This leaves the analyst with a very precise picture of the outcome of one particular battle fought in a few particular ways with certain pieces of equipment modelled to some approximation of reality. Due to the "potted" nature of the set-up process, one is often left with the impression of having been able to predict the outcome of the run before it happens.

On the other hand, the ability of models like ISAAC to adapt to the situation as it develops during the run explores a much broader range of possibilities. By doing so, there is potential to gain information which the analyst had not anticipated beforehand.

Here we will give just a brief overview of the workings of the ISAAC model. The reader is referred to Ilachinski (1997) for more detail. Additionally, the model itself may be downloaded from CNA's Website (www.cna.org/isaac), which also contains information on the workings and philosophy of the model.

The behaviour of the automata is determined by local rules, that is, they react to the situation within their sensor range.

The rules are predominantly determined by three sets of parameters. The first set describes weapons capabilities. These are the weapons range, sensor range, movement rate, single-shot kill probability, defensive factor and maximum number of simultaneous targets which can be engaged.

The second set is the "personality" parameters, which determine the relative weightings with which the automatons are compelled to move towards the opposition's flag/their own forces/the opposition forces.

The final set are the "meta-personality" parameters, which determine how many friendly automatons a given automaton must have with it before it will advance towards opposing elements, the maximum cluster size of friendly forces, and the minimum local numerical advantage the automata of a particular force must have before they will engage in combat.

2.1 "Napoleonic" squares

This section and those following explore how increasing degrees of complexity in manoeuvre affect combat outcomes. Here, a brief description is given of each manoeuvre case studied. In this first example, a simple attrition-based situation is examined. The situation is linear in a military sense (i.e. the two forces form a nice Euclidean geometric shape), and may be reasonably closely described using a linear differential equation (leading to the Lanchester square law), since the two sides simply "shoot it out" with no manoeuvre. The situation is somewhat reminiscent of Napoleonic warfare (or perhaps more similar to the tactics of Fredrick the Great's Prussian army), where we might imagine that some manoeuvre has occurred to bring the two squares together before hand. The forces would typically be assumed to withdraw once a certain level of attrition is reached.

While such a case is reminiscent of the linear methods of warfighting (where opponents formed lines to concentrate firepower), it lacks the human element. Napoleonic warfare was, of course, more complex than the case presented here. In particular, factors like troop quality and morale were extremely important. The intangible nature of these factors generally means that they are not modelled by conventional analytical wargames. Typically, such conventional models manoeuvre the opposing forces into each other's range, then fight to some level of attrition, where upon they manoeuvre to some new position. Thus such a conventional model may be viewed as being very similar to a series of fights of the kind discussed in this subsection, even though the situation being described may not be intended to be an example of "linear" warfare.

The ISAAC run shown in Figure 1 is an example of a model whose outcome is dependent entirely on kill probabilities, the winner determined by who remains at the end.

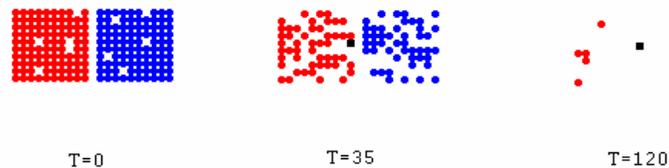

T=0          T=35          T=120

Figure 1: Determining combat without manoeuvre. Though a very simple situation, it should be noted that conventional combat models tend to be little more than a series of such fights.

The ISAAC parameters can be found in the appendix under the name "Fluid_0D". An important point to note is that the behaviour of the model remains the same regardless of the starting number of automata. This may seem an obvious statement, but the same will not be true of the examples shown in the following subsections. As such, this is an important difference between adaptive combat models such as ISAAC and conventional combat models.

## 2.2 Fluid battle

For the next simulation, we allow the automata to manoeuvre according to the ISAAC rules. We do this for two cases, the first with 95 automata on each side, the second with 10 automata on each side (note that both sides have the same parameter set). The parameters used are given in the appendix under "Fluid_1D".

Here each force starts in an opposing corner and attempts to capture the flag in the centre. Figure 2 shows a typical evolution of the battle. At time $T = 20$ the two forces approach the flag (black square in middle of battlefield); $T = 50$ sees the two meet, by this stage both forces have formed a line as they fight for the flag; $T = 100$ sees the red force pushing the centre of the blue line away from the flag; $T = 200$ sees a continuation of this trend, though the blue line has not yet broken; $T = 400$ sees the blue line broken, yet surprisingly there are more blue remaining than red; $T = 795$ sees red still in control of the flag, with the blue force driven off to the fringes.

The interesting thing to note about this battle is how linear it is up to its end state. The formation of opposing lines appears to dominate the dynamics of the battle, and this line formation survives in the case shown until the point where the blue forces are decisively dispersed by the blue onslaught. Interestingly, such an end state does not occur in conventional models without significant artificial interference from the modeller.

The nature of this battle changes markedly when the starting number of automata is reduced to 10 on each side. The evolution of a typical case is shown in Figure 3. Here, at $T = 35$ the two forces approach the flag. Note that the red force is broken in half. This behaviour actually only occurs on the minority of occasions, but the tendency of a small group of automata to split into still smaller groups is significant, since such a split tends to represent a large portion of the force. This contrasts with the case where each force had 95 automata, since splits from the main force were small relative to its size.

At $T = 60$, the forces clash (note that the illustration is cropped so that the flag is at the top right-hand corner where the large cluster of automata is). By $T = 80$ the blue force has captured the flag (not surprising, given their local numerical superiority). The red force, however, is not defeated, and remains around the fringes picking off stray blue automatons.

This remains the case until around $T = 400$, when suddenly the blue chase aggressively after red. This change in behaviour was brought about by the loss or withdrawal of one or several of the red automata surrounding the blue. The next phase of the battle sees the blue and red forces rapidly moving across the battlefield in a "chaotic" manner. This brings into play the remaining red forces (who by this stage, still had not moved close enough to join the combat. Note that an ability to communicate between the two groups of red would have drawn the stragglers in earlier). The resulting clashes between the red and blue even up the battle, so that by time $T = 1000$, all the blue are injured, while all red but one are injured. Now, neither side is capable of holding the flag. Interestingly, the splitting of the red force has ultimately not disadvantaged it in this run.

The key difference between the cases shown in Figure 2 and Figure 3 is that for the former case, attrition occurred relatively constantly through the course of the battle, while in the latter case, the battle was marked by a series of decisive clashes which tended to change the course of the battle drastically.

Thus the dynamics of the latter case appear to be more complex than for the former. This suggests that the latter has more degrees of freedom. At first sight, this suggestion does not seem intuitively obvious, because the former case utilised a much larger number of automata. However, the reason why the case with the larger number of automata has fewer degrees of freedom is that it has more constraints. That is, a large number of automata tend to constrain themselves to fight as a single group. In this sense, the automata display mob behaviour. Above a certain size for the group (in this case, 15 or so automata), this behaviour does not seem to alter. For a small group of automata, however, the dynamics depend quite strongly on the proximity of friends or enemy, particularly when the size of the group is on a par with the parameters which determine such things as likelihood to advance or cluster. The upshot of this is that small groups tend to split more readily, and generally act as individuals.

The effect of this behaviour on the statistics will be discussed in section 6, where a more quantitative aspect will be added to the qualitative observations presented here.

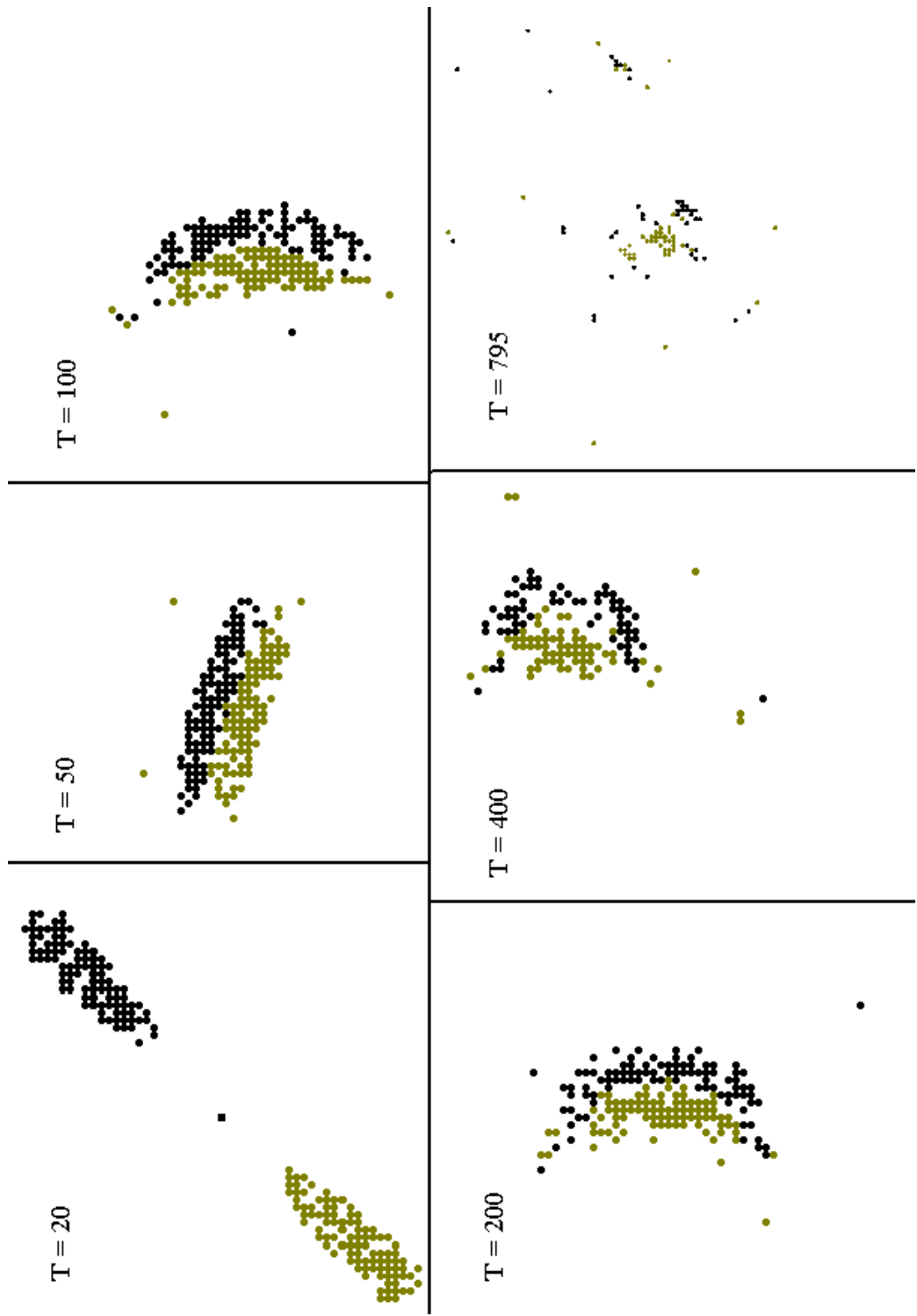

Figure 2: Evolution of the battle between two forces of 95 automata using the

Fluid_1D parameter set.

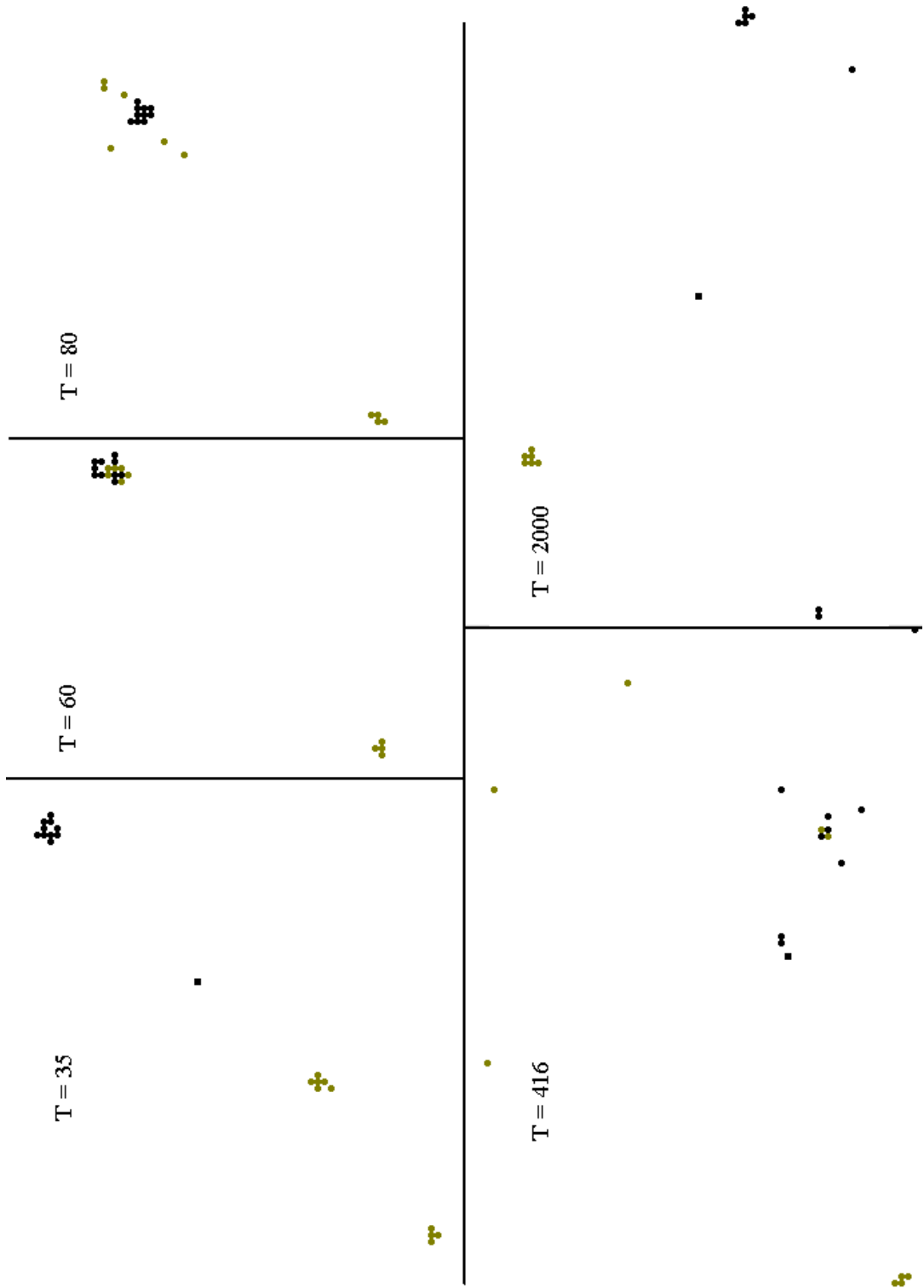

Figure 3: Evolution using the Fluid_1D parameters, but with 10 automata on each side.

2.3 Effect of additional unit types

The next study considers the effect of adding more automata types to the battle. This increases the complexity. Here, in addition to the automaton parameters used in section 2.2 (notionally describing infantry), we introduce two new sets of parameters describing different element types.

The first additional set describes notional artillery. While once again we present the various parameter values in the appendix (see Fluid_3D), we give a qualitative description here of the capabilities of this type. Automata acting as artillery are slow-moving, have long-range weapons with high kill probability, and tend to stay as far from enemy automata as possible. In order for the artillery units to make use of their range, they have a long sensor range (thus notionally possess forward spotting information).

The second type is a notional tank unit. These are fast, with medium range and high kill probability. Once again, to make use of their range they tend to stand off somewhat from the enemy.

Figure 4 shows the evolution of such a battle. At $T = 4$, the tanks from both forces are moving ahead of the main force towards the flag (centre). By $T = 15$ the two tanks forces have reached the flags and are engaging each other, with the infantry and the artillery bringing up the rear. At $T = 33$ the infantry has arrived, and the battle formations have striated out, with infantry at the front, tanks behind and artillery at the rear. At $T = 80$, the red force has pushed the blue force away from the flag; however during the next period the blue force is able to regroup and counterattack so that by $T = 150$ the blue force is able to push the red force away and take control of the flag.

The dynamics of this simulation bear a stronger resemblance to the case shown in Figure 3 than that in Figure 2. That is, the battle consists of a series of decisive clashes (the tank encounter, the arrival of the infantry, pushing back of the blue forces and the blue counter attack), the outcome of each changing the evolution of the battle markedly. This was despite the run having a similar number of automata to the case shown in Figure 2.

The reason for the similarity in the dynamics of these two cases lies with the presence of three distinct automata types on each side. The dynamics are such that each force breaks up much more readily than for the case shown in Figure 2, and these groups themselves behave as if individual automata. Thus instead of each force being constrained to act as a single entity, as was the case with the run in Figure 2, each force acts as a collection of entities, with analogous dynamics to a small collection of automata as shown in Figure 3.

Thus, unlike the case discussed in section 2.2, the dynamics show a self-similarity as the size of the force involved increases.

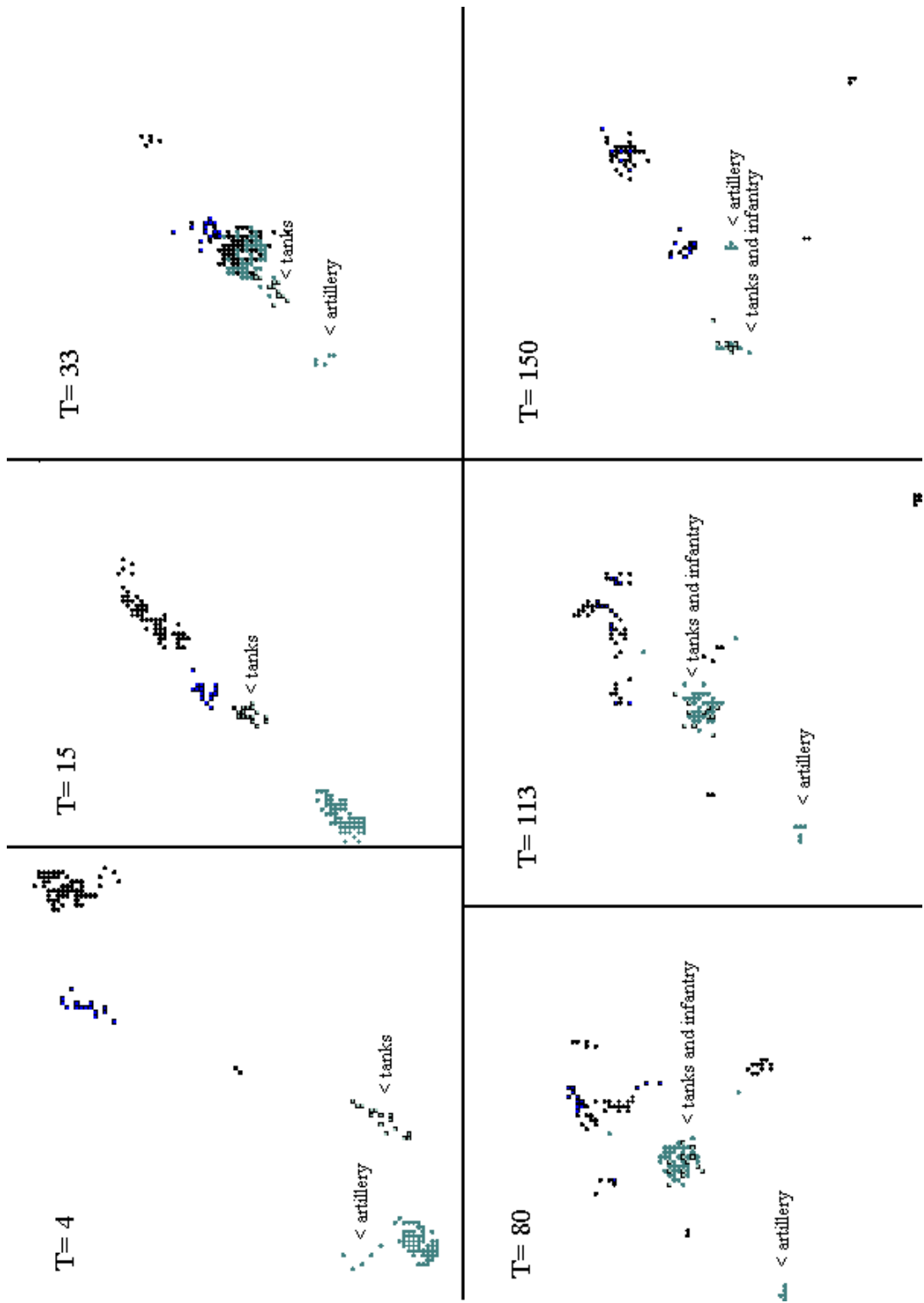

Figure 4: Evolution of a battle using the Fluid_3D parameter sets.

### 3 Statistics of complex systems versus attrition-driven models

An important theme in this work is to be able to determine that complexity, and more specifically, complexity brought about by manoeuvre, affects the statistics of warfare.

One traditional approach to modelling combat mathematically is to assign each side an attrition rate, and use this to calculate the number of casualties after a certain length of time. Such a model may be represented as a pair of coupled ordinary differential equations, where if the attrition rate is linearly dependent on the number of participants. This leads to the Lanchester square law model. For the purposes of this report, we refer to such a model as "attrition driven" (since attrition rate is the only parameter).

More recent (but still "conventional") models make use of computer power to model the individual elements in a given battle. While these elements can manoeuvre in any way desired, generally the models will be characterised by periods where the forces manoeuvre into position for combat, and then fight an attrition-based battle to some arbitrary point of the modeller's determination, where upon they withdraw (or take some next step). That is, the model is really a series of attrition-driven fights, with the manoeuvre serving to determine who is involved in those fights. In order to get a statistical spread of results, such models are run many times (though still usually only of the order of tens of runs). However, in the majority of these models, the variability is almost entirely due to the modelling of hit and kill probabilities, with little or no recognition of the way battlefield entities react to the changing situation i.e. modelling of "command" is poor or non-existent.

Where this approach differs from those researchers who are proponents of using complexity theory to model the battlefield, is that these conventional models do not incorporate a key human trait — that the participants adapt to the situation as it evolves. By contrast, the ISAAC automata manoeuvre within each run so that they will only attack when conditions are favourable to them, determined by their "personality" parameters.

For any model which allows manoeuvring, one would expect phases of manoeuvre (with little or no attrition), punctuated by periods of fighting. This being the case, it may be claimed that manoeuvre combat can be described by an attrition-driven model, where the attrition rate is determined by a combination of the kill probabilities and mean time between shots. This is certainly the case for conventional combat models, because the time spent manoeuvring between fighting will vary little from run to run (in fact, for most such models, it doesn't vary).

However, for ISAAC the time spent manoeuvring does vary between runs. But we may still apply the claim above by using the mean time spent manoeuvring. However, such an endeavour is not as straightforward as it might appear.

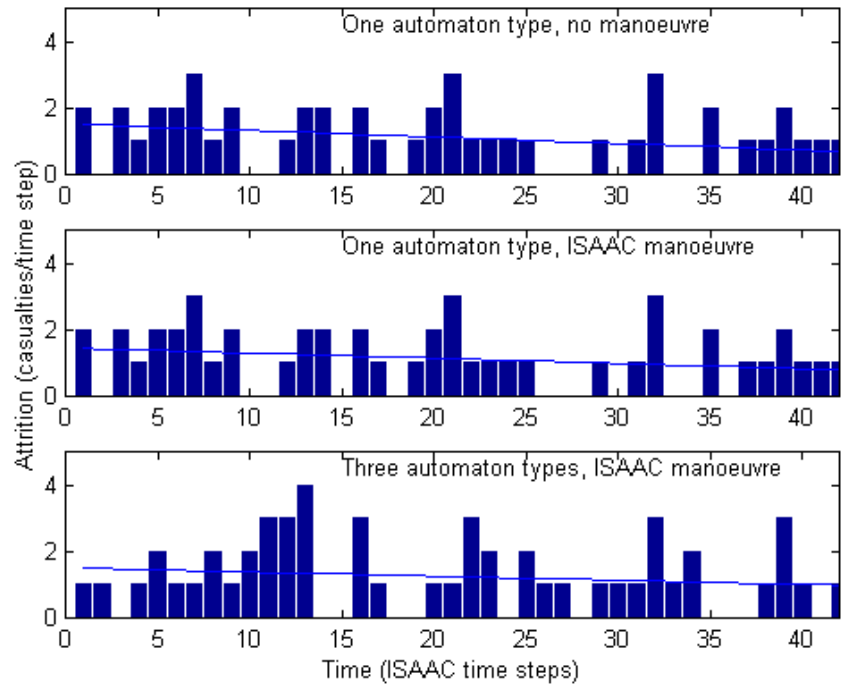

Figure 5: The attrition function found using three variations on ISAAC parameters.

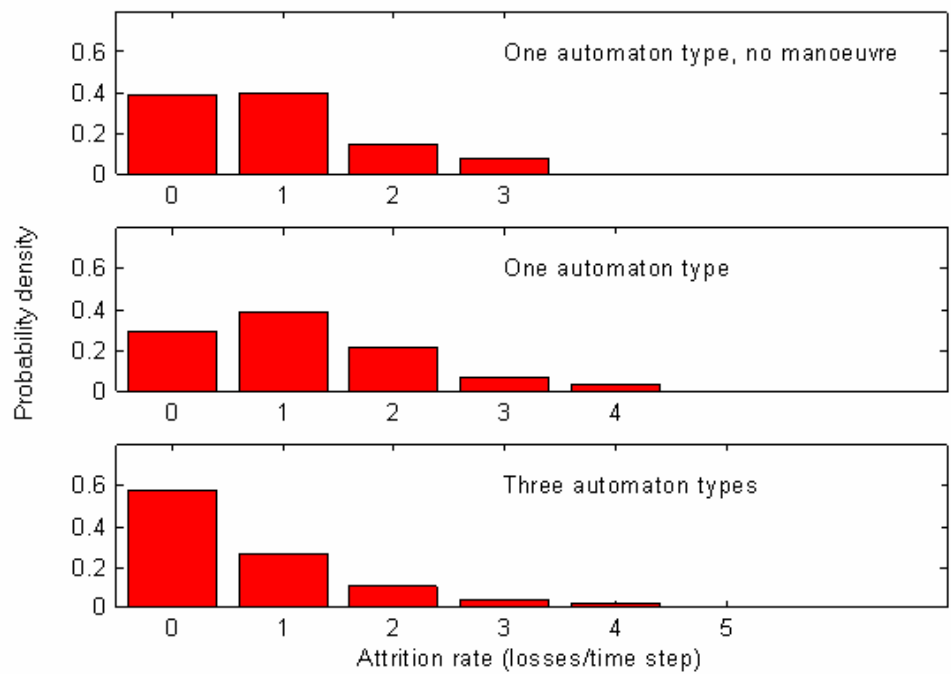

Figure 6: Probability densities for the data shown in Figure 5.

Let us consider attrition data generated by ISAAC in Figure 5. This shows three runs of the ISAAC model, each with a different level of complexity. These correspond to the cases discussed in sections 2.1 (top), 2.2 (middle) and 2.3 (bottom), with 95 automata on each side.

The data shows attrition to the point where casualties reach 50 per cent. At a glance, it is not obvious that the data presented in each of the cases is fundamentally different from the others. However, while it is possible that any given set of data from a single run from the non-complex case may look similar to a complex case, and even possess a similar-looking probability distribution, this will not be the case as the ensemble of runs grows.

Figure 6 shows the probability distribution for attrition data points (that is, the value of attrition at any given instant of time) from six runs of each case, giving a total of 200-300 data points each (note that the data is taken from the time when the opposing forces first meet, to eliminate the zero values for attrition leading up to this point). From the figure it appears the distributions have a longer tail the more complex the model run is. This is consistent with the data being more intermittent as the complexity increases, since intermittency increases the probability of zero attrition values, and consequently requires a higher probability of large values for attrition to maintain the same attrition rate.

What is more surprising, and not obvious from the figures presented so far, is that even random data generated using the distribution shown at the bottom of Figure 6 will not reproduce all the characteristics of ISAAC data (that is, "synthetic" ISAAC data cannot be produced solely from a probability distribution generated from a given ensemble of runs).

There are two reasons. The first is that the attrition data in the bottom case of Figure 5 possesses temporal correlations (i.e. the data cannot be treated as a series of independent random variables). We leave the examination of this property for the next section.

The second is that trying to reproduce ISAAC data in this way depends on both the ISAAC data and the synthetic data reaching the 50 per cent casualty level in roughly the same period of time (since they both should have the same mean attrition rate). Although the time taken to reach this point in both cases varies (due to the random element in the models), we expect the variation to have a well-defined distribution. However, this is not so for actual ISAAC data.

Figure 7 explores this. Here, two distributions of the times taken to reach 50 per cent casualties for the red side are shown. Each case modelled combat between two identical forces each with 10 identical automata. The first (circles) had no manoeuvre occurring (the attrition-driven case). The second case allowed manoeuvre occurring to the ISAAC rules. The plot shows the distribution for 1000 runs of the first case, and 200 runs of the second.

The point to note from the second distribution is the extreme values in the "tail" of the distribution. The figure shows the distribution up to 800 time steps. Additionally, there were four values lying between 1000 and 2000 time steps, and one value of 9200. These were not plotted to prevent the plot becoming too condensed to read. Also, there were five occasions when 50 per cent red casualties were never reached (because all the blue side had been killed). On the other hand, for the first distribution, the maximum time was 255, and of the 1000 runs, only once were all the blue side killed before 50 per cent red casualties were reached (i.e. this was 25 times more likely to happen for the adaptive rules).

This behaviour causes the variance of the distribution to diverge (that is, for the number of runs used, the variance doesn't settle down to a fixed value). This is illustrated in Figure 8. The divergence comes about because as more runs are examined, ever more extreme cases emerge. This behaviour suggests that these extreme values actually belong to separate statistical populations, which themselves belong to different states into which the system may evolve.

By a "state" it is meant that the model exhibits behaviour which is qualitatively the same and statistics (such as mean attrition rate, the fractal dimension of the attrition function, and the distribution of possible attrition values) which are quantitatively the same for each such state.

So for the complex case, the ISAAC model may evolve into different states with significantly differing statistics. In the states which correspond to the extreme values observed, the attrition rate becomes significantly lower than a "typical" run.

Furthermore, it is extremely difficult to predetermine which states the system will evolve into (in fact, where there are random factors in such models, it is impossible). This is a key different between the ISAAC model and conventional combat models.

Qualitative examination of the model run reveals that the extreme values arise as a result of the adaptive nature of the ISAAC rules, which cause the automata to avoid combat unless conditions are suitable. In some circumstances, the automata will manoeuvre for an extended period after the initial contact (having suffered significant casualties) before finding conditions suitable to attack again.

The existence of multiple states into which the model may evolve is particularly interesting for analysing low-risk situations, where standard analysis methods might not reveal potential extreme circumstances where a situation suddenly becomes much more dangerous. Alternately, high-risk situations may be able to be negotiated with low casualties in certain extreme circumstances. It is thus suggested that adaptive analysis methods should be used to explore the possibility of extreme outcomes, and that these extreme cases contain information that could guide the analyst to find where there is potential to substantially improve performance.

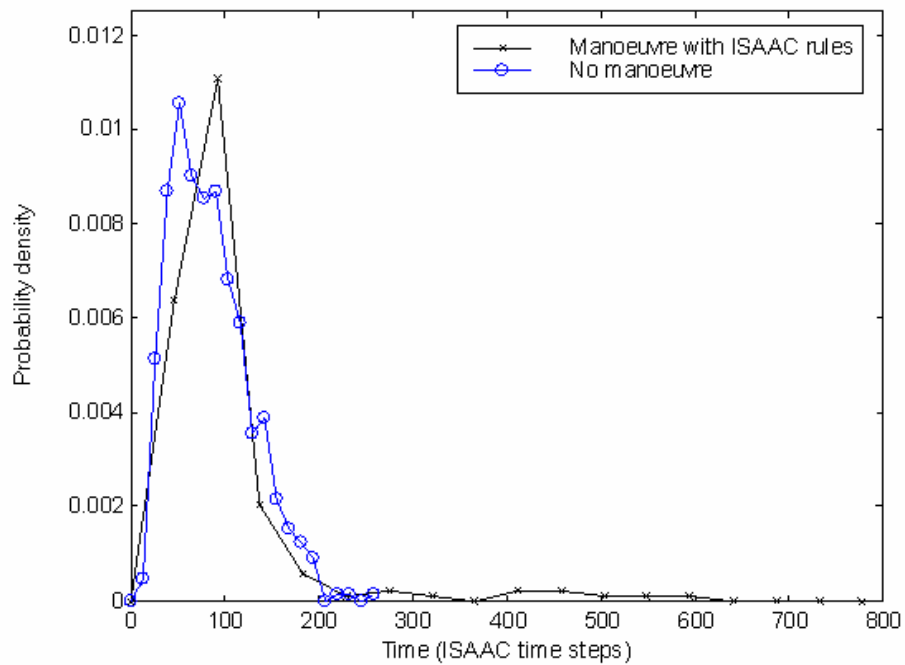

Figure 7: Probability densities for times to reach 50 per cent casualties.

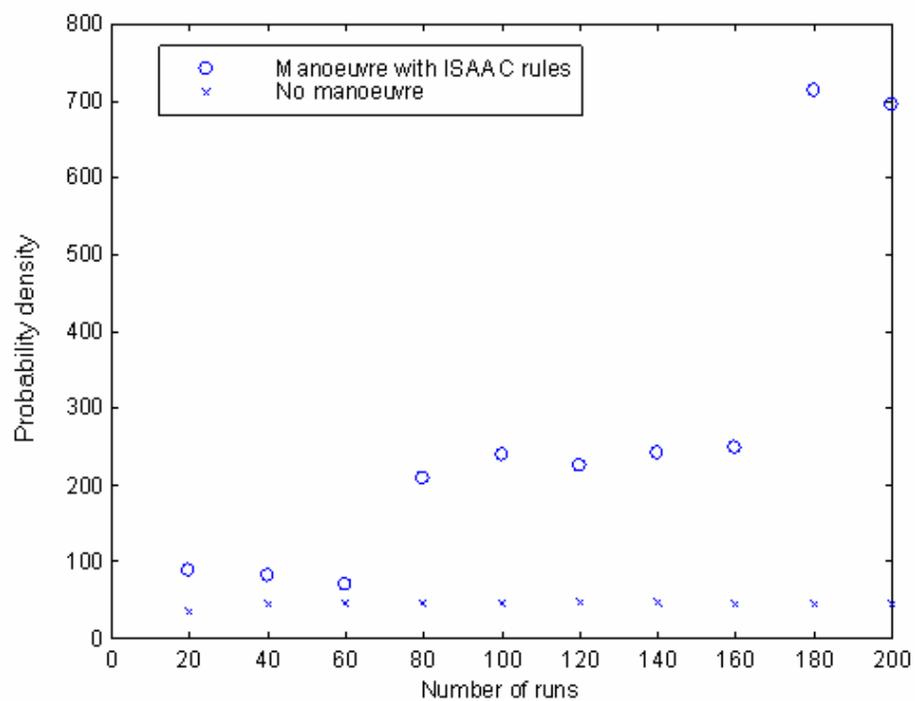

Figure 8: Standard deviation as a function of the number of runs examined.

# 4 Application of fractal methods to ISAAC statistics

Intuitively, one expects attrition data in real warfare to demonstrate temporal correlations. In reality, battles are not a series of random events — they possess structure. On the battlefield, neither side's troops constantly shoot, but instead manoeuvre in an attempt to gain some advantage over the enemy (an action which leads to intermittency of casualties). When such an advantage is perceived to have been gained, the fighting begins in earnest.

The difficulty in modelling this is that in the heat of battle any imposed structure (from higher command) is likely to break down. A different kind of structure emerges, that is, a self-organised structure. The ability of troops to organise themselves well enough to fight under trying circumstances is something which is extremely difficult to describe. It incorporates many of the intangibles of war, such as troop quality, morale, initiative and adaptability.

Models such as ISAAC are the first attempt to measure these things. However, measuring ISAAC itself may prove just as difficult a problem. Fortunately with ISAAC, we have the necessary reproducibility to make quantitative statements about the behaviour. We may thus seek to describe the sort of battlefield structure which often seems to emerge "by itself" in ISAAC. Further, ISAAC data displays temporal correlations.

In this section, we introduce methods for characterising such properties. Principally this is done by the use of fractal methods. These methods, as has been pointed out by Mandelbrot (1983), are ideal for characterising data which are autocorrelated, clustered and possess fat-tailed probability distributions.

## 4.1 Smoothness of data and fractal dimension

Consider the data presented in Figure 5 in the preceding section, which shows attrition as a function of time for ISAAC model runs up to the point in time where the casualty level reaches 50 per cent.

Clearly, the attrition function is not well behaved, being neither smooth, nor non-zero. Such a function is not easily modelled by a differential equation. Also, as already stated, there is no qualitative difference between each case which would allow us to draw any conclusion from each dataset. However, the use of fractal analysis can quantitatively determine the difference between each set of data.

One such method is to find the fractal dimension of the attrition function for each case. This can be done using a so-called box-counting technique (see Mandelbrot 1983). That is, imagine a box one attrition unit high and one time step wide, and count how many are required to cover each of the functions shown in Figure 5. Then, increase the height and width of the box, say, 10 per cent, and count how many are required again, and so on. If a fractal dimension exists, then the plot of the number of boxes required versus box size should produce a straight line on a log-log plot. Figure 9 demonstrates this procedure.

Figure 10 shows the plot of box-width versus number of boxes required. The slopes for each case (from top to bottom, using a linear regression) were −0.87, −0.92, and −0.78 respectively. The difference in these values is interpreted as describing the differing degrees to which each function in figure 2 displays scaling structure (that is, structure on many scales). The results obtained suggest the last case possesses more scaling structure than the first two. Note that for a continuous attrition rate (and thus a perfectly smooth function), the slope is −1.0, as represented by the solid line in Figure 10. This parameter (and those following) were determined from ensembles of data from six different runs of each parameter set, so that in fact 200-300 data points were used.

Generally, a large number of runs need to be analysed in this way to give an accurate estimate of the fractal dimension (see Lo *et al.* 1993, for an example). The slopes found above determine the fractal dimension $D$ of the attrition function, defined as:

$$D = \lim_{d \to 0} \frac{\log N}{\log\left(\frac{1}{d}\right)} \tag{4.1}$$

where $d$ is the width of the box, and $N$ the number of boxes required. If $D$ is a non-integer, then the attrition function is a fractal. If this is so, one expects the number of boxes required to go as:

$$N(d) \sim d^{-D} \tag{4.2}$$

We will see in section 5 that the existence of a non-integer fractal dimension implies that the spectrum of the data obeys a power law; that is, the data is not "white" random noise, but exhibits temporal correlations.

Figure 11, Figure 12, and Figure 13 demonstrate how this affects the statistical characteristics of the three attrition functions considered above. These figures show the probability density for attrition values conditional on the preceding value being above a certain level. Of course, if the data is independently random, then the preceding value makes no difference to the distribution. If the distribution does depend on the preceding value, then there is temporal correlation.

From the figures, it is clear that the distribution changes as the conditional value increases (compare these figures with Figure 6, produced from the same data). In particular, for Figure 13 there is a much greater likelihood of obtaining an attrition value of greater than 2 when the preceding value is also greater than 2. In other words, when it rains, it pours. Such properties may be crucial in contingency planning.

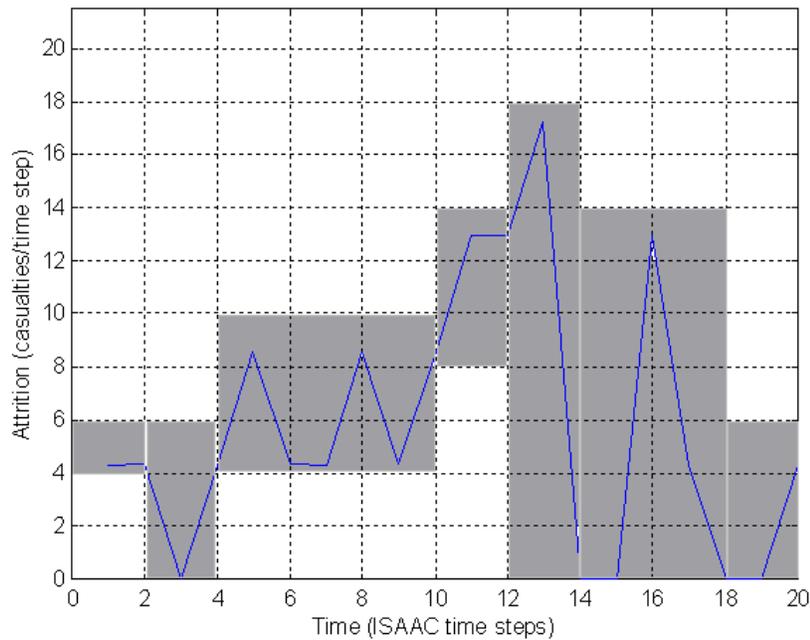

Figure 9: The attrition function here lies in a 20 x 20 square. Divide the square into 100 boxes of width $d = 2$, and count the number of boxes $N(d)$ in which the function

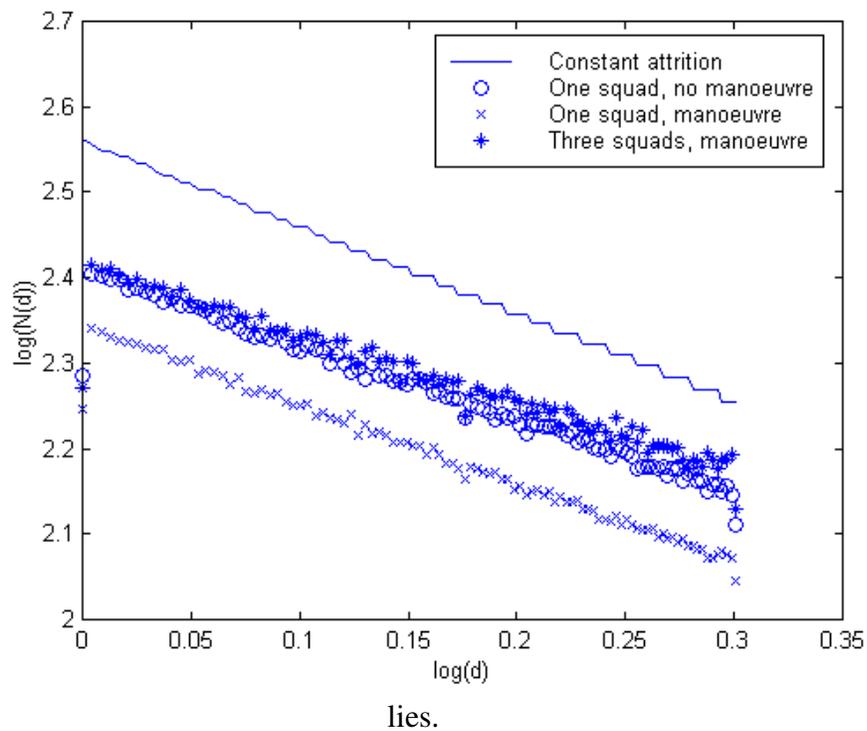

lies.

Figure 10: Number of boxes $N(d)$ required as a function of box width $d$.

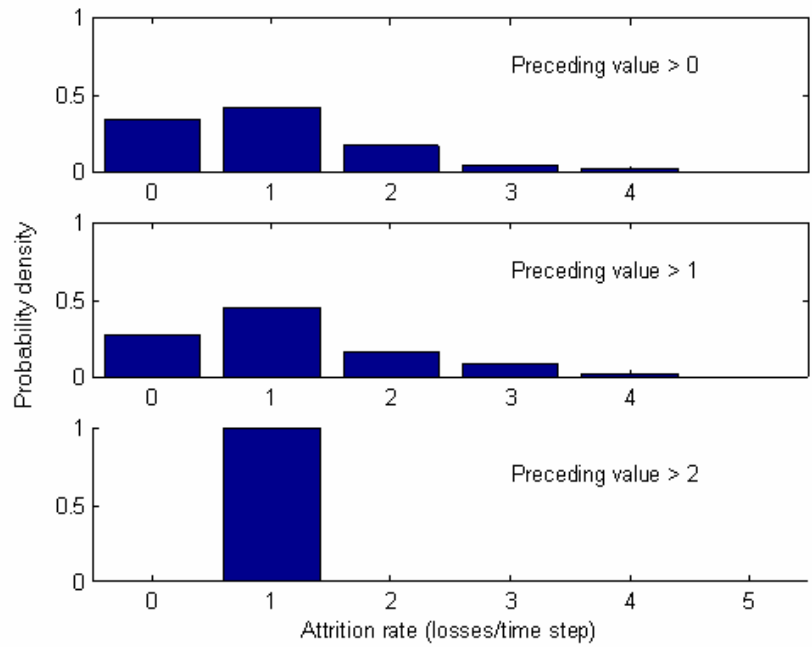

Figure 11: Probability density conditional on the preceding value being greater than 0, 1 and 2, for the attrition function with a fractal dimension of $D = 0.13$.

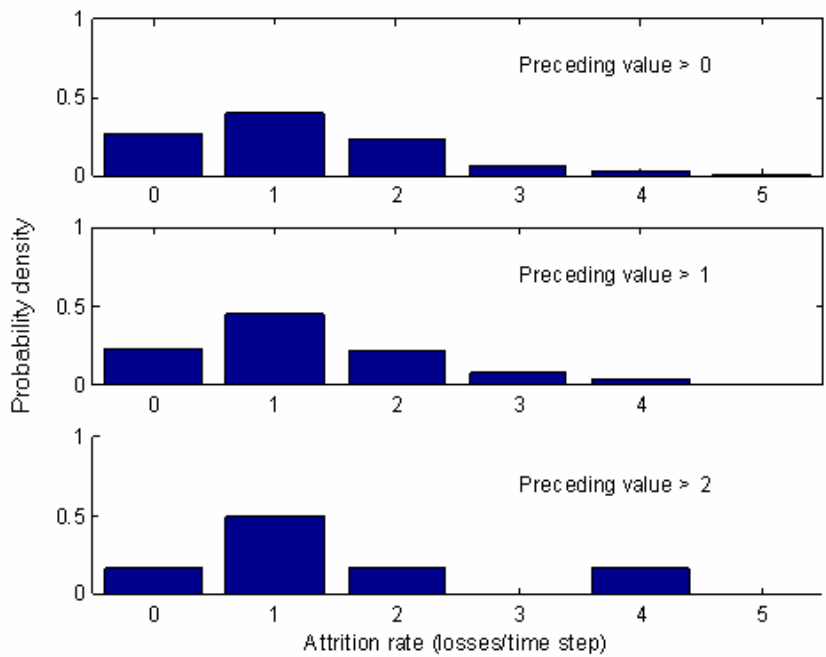

Figure 12: Same as Figure 11, for the attrition function with $D = 0.08$.

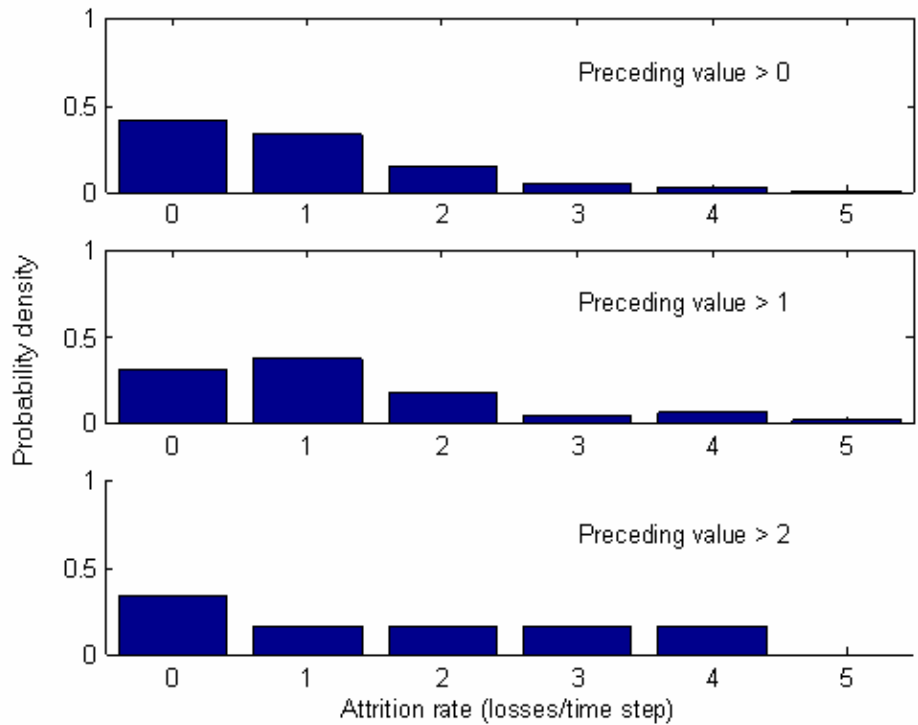

Figure 13: Same as Figure 11, for function with $D = 0.22$.

## 4.2 Measuring intermittency and the usefulness of fractal cascade theory

The fractal dimensions obtained above represent the degree of "structure" within the data. This implies that, to an extent, the small-scale structure of the battlefield resembles the large scale. As has been noted, on the turbulent battlefield the forces engage each other by forming clusters. The fractal dimension describes to what degree a large cluster can itself be viewed as a collection of smaller clusters, which themselves are collections of even smaller clusters, and so on. Such an arrangement describes a "mono" fractal (i.e. it is described by a single fractal index).

The property of intermittency breaks this simple scaling, so that one index is not sufficient, but there is a continuous spectrum of fractal indices. This idea is reflected in the multifractal formalism (a good description of this is in Davis *et al.*, 1994). Intermittency can be seen on the turbulent battlefield by the existence of large empty patches. That is to say that, although the turbulent battlefield does not possess nice linear formations, neither is it filled with a uniformly random spread of combatants. There are large empty patches where nothing is happening, but smaller regions where there is intense activity.

This is reflected in the attrition function, in that casualties are not constantly occurring, but rather come in bursts. For attrition data of this nature, the statistical moments obey a law:

$$\left\langle a_i{}^q \right\rangle \propto \left( \frac{t}{T} \right)^{-K(q)}$$

where $a$ is the attrition rate at the $i$th time step, the angled brackets represent an ensemble average, $t$ is the temporal resolution at which the distribution is being examined, $T$ is the "outer" scale of the scaling range, and $K(q)$ is a non-linear function of the order of the statistical moments, $q$.

This equation implies that the statistical moments of the data scale. Figure 14 demonstrates the scaling of the second-order moment (variance) for the data in Figure 5. Fitting straight lines to the points in Figure 14 provides the value of $K(2)$ for each case. The slope tends to be steeper the more intermittent the data. Thus a popular measure of intermittency is the parameter $C_1$, defined as the derivative of $K(q)$ at $q = 1$. Figure 15 plots the $K(q)$ function for each case. Note that the function is non-linear, as expected for intermittent data.

The intermittency parameters for each case (once again, from top to bottom of the data in Figure 5) are 0.053, 0.037, and 0.070. Note that the values given were the average from six different runs, and so are estimates based on about 300 data points. While this is not a particularly large sample, so that there is at least a 15-20% uncertainty associated with each (see Harris *et al.* 1997), it is probably sufficient for illustrative purposes.

Though this method is not the only one capable of measuring intermittency, it has the appeal that it is closely related to fractal cascade models. Cascade models provide a stochastic means by which "artificial" ISAAC data may be generated, and consequently present a method for completely characterising the data obtained from ISAAC.

If ISAAC data is suitable for characterisation using these methods, then use of the known properties of cascade models can be exploited to approximate the statistical behaviour of ISAAC data for various cases. It must be noted that there is no physical justification for the use of a cascade model in this case. Rather, it is employed here as a statistical method. This also appears to be the case where these methods have been employed to describe financial market statistics (e.g. Mandelbrot, 1997), since it is clear that such markets have no underlying cascade process. At any rate, without such a framework it is virtually impossible to derive analytically the statistical character of complex models such as ISAAC.

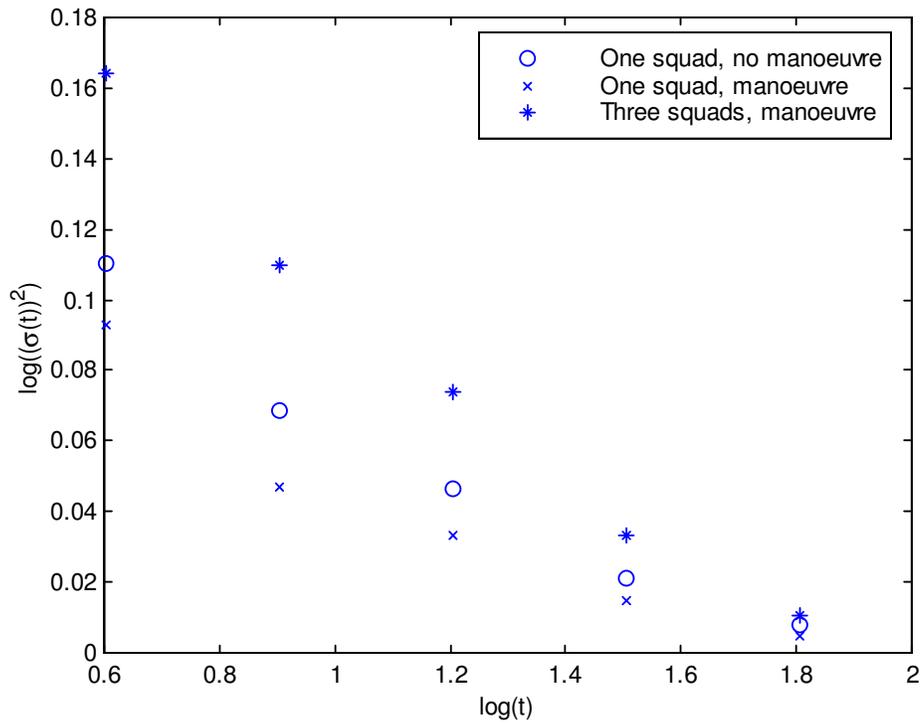

Figure 14: Scaling of the second-order moments for the data from Figure 5.

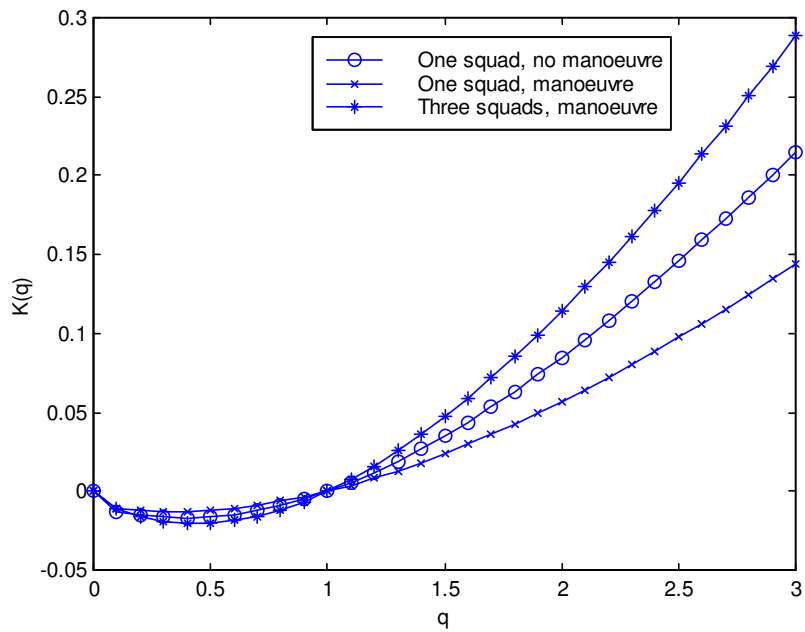

Figure 15: The $K(q)$ functions for the data in Figure 5.

Having obtained these and the other fractal parameters, it is in principle possible to simulate ISAAC data representing each of these cases via a fractal cascade. This may be especially useful for characterising the extent to which the data from each case exhibits fat probability tails (as will be discussed in the following sections), because the degree of intermittency is related to the likelihood of extreme events. Additionally, understanding the nature of the "burstiness" of the data may be useful for force maintenance modelling, particularly if the time scales being considered are large enough to allow pools of replacements time to be brought in.

Most importantly, obtaining these parameters provides a method for characterising the degree to which a given model has the capacity to exhibit complex behaviour.

## 5 Attractors for the ISAAC model and the form of the statistical structure function

At first sight, the ISAAC model does not appear to be a "useful" physical model because of the large number of parameters, from which it is virtually impossible to derive analytically the statistical behaviour.

However, the model may still be interesting if it exhibits behaviour that is not strongly dependent on the values of these parameters, but instead appears to behave in qualitatively similar ways for a wide range of parameters.

Also, since the model may exhibit complexity, behaviour may be sought which is qualitatively independent of the initial conditions. Such behaviour is a sign that the model has evolved into an attractive state for the system. If the attractor is peculiar to the model, characterising it may be more useful than trying to analyse the results of any given run, or ensemble of runs.

Consider a set of parameters for ISAAC called "Einstein_fluid" (available from the CNA Website), given in the appendix.

This set of parameters is so named because it is supposed to represent a simulated engagement that evolves as though it were a clash between two "viscous fluids." Figure 16 shows how the automaton battle progresses. At T=13 the opposing forces move towards each other. At T=26, the forces have made contact, forming roughly a line. However, the line formation is unstable, and has begun to break up at T=50. By T=102, the battle no longer has form (in that it has no traditional military formation).

Though it is difficult to describe the distribution at T=102 in terms of Euclidean geometry, it may be described by fractal geometry (just as other cellular automaton models can, as discussed by Bak *et al* 1989, 1990). This can be quantified by use of a renormalisation procedure.

Figure 17 shows such a fractal distribution of automatons. To see that the distribution is fractal, the "battlefield" square is split in four, and the number of squares containing automata counted. Then each of these new squares split in four, with automaton-containing squares counted again, and so on. If the distribution is fractal, then the proportion of the number of squares containing automata should show a power-law dependence on the size of the squares used (this is a box-counting technique, see Mandelbrot, 1983). Figure 18 suggests that this is the case over the range concerned.

Let us change this simulation in the following way: the size of the battlefield is increased from 80 x 80 to 120 x 120, and the personality weighting which determines how compelled the red elements are to move towards the blue flag is set to zero, while for the blue elements this is set to 1.

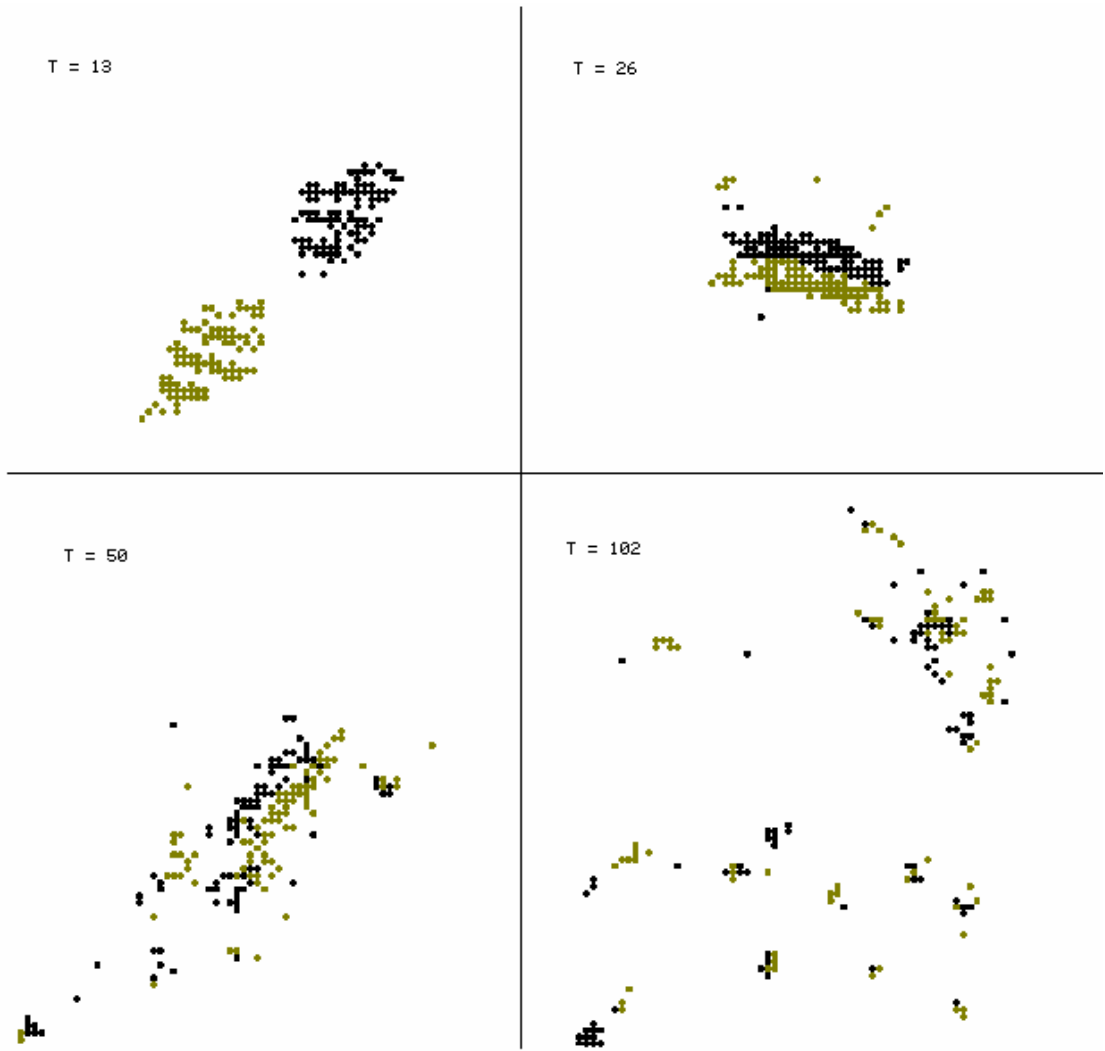

Figure 16: Evolution of the "Einstein_Fluid" battle.

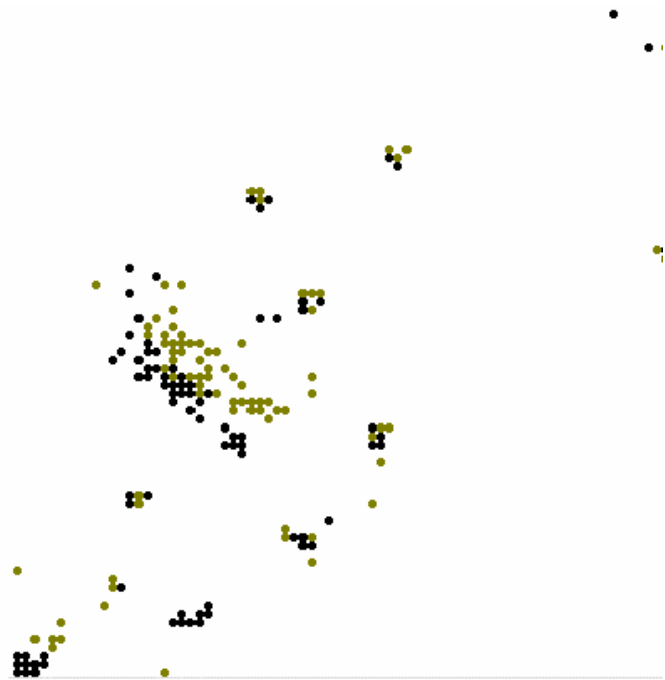

Figure 17: Snap-shot of "Einstein_Fluid" simulation used for renormalisation analysis.

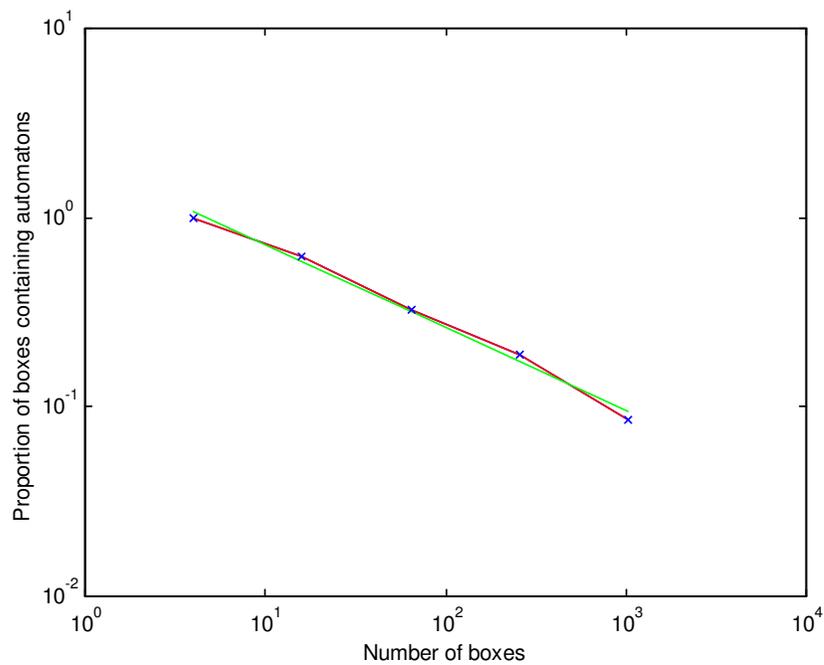

Figure 18: The renormalisation analysis shows the number of boxes required depends on box size, and obeys a power law (which is related to the fractal dimension).

As a consequence, the opposing forces concentrate on responding to each other rather than rushing for the opposition flag. Figure 19 demonstrates such a run (the parameters are given in the appendix under the name "Transition").

At T=50, the blue force is proceeding towards the red flag, while the red force waits to block the advance. At T=100, the forces have established contact, and begin to form two opposing lines, as with the "fluid" simulation. However, the difference this time is that the line formation appears quasi stable, as seen at T=150, and remains so for at least another 50 time steps (even though there is some breakdown around the fringes of the opposing lines, as seen at T=200).

At some point, which is difficult to determine precisely, the line formation rapidly breaks down, and the distribution of the forces across the battlefront quickly turns from a line into the same fractal-like distributions that the "Fluid" parameters generated. Thus there are two states in the battle, a "linear" phase, and a "fractal" phase.

In the initial phase of the simulation, the two forces have the same number of elements, so that neither is able to force the other back. The two forces spread out to form a line, which is relatively stable. Thus this two-line formation is one attractor for the dynamics. Note that the simulation does not strictly form two lines, but fluctuates about the attractor. As the simulation proceeds, the formation twists and turns to maintain this situation. Eventually, some small perturbation leads to the collapse of the entire formation. The dynamics are then attracted towards a fractal distribution.

Qualitative observations of the model runs suggests two factors determine the evolution of the model:

• The model reacts to reduce imbalances in the local number of red and blue forces (notionally force "gradients").

• The degree to which each force clusters together (notionally friction).

Further, the much larger number of model parameters appear to set the levels for which such force imbalances and clusters are important. As such. we may consider the two factors pseudo parameters. Thus, the local balance of the forces drives the model, rather than attrition which, as discussed in the preceding section, is the main determiner of manoeuvre for conventional combat models.

The two types of formation which serve best to minimise force gradients in this case appear to be a line and a fractal distribution. Hence the model evolves to one of these two states. In this sense, the behaviour of the model is determined by geometry rather than the parameters (within some range of parameter values).

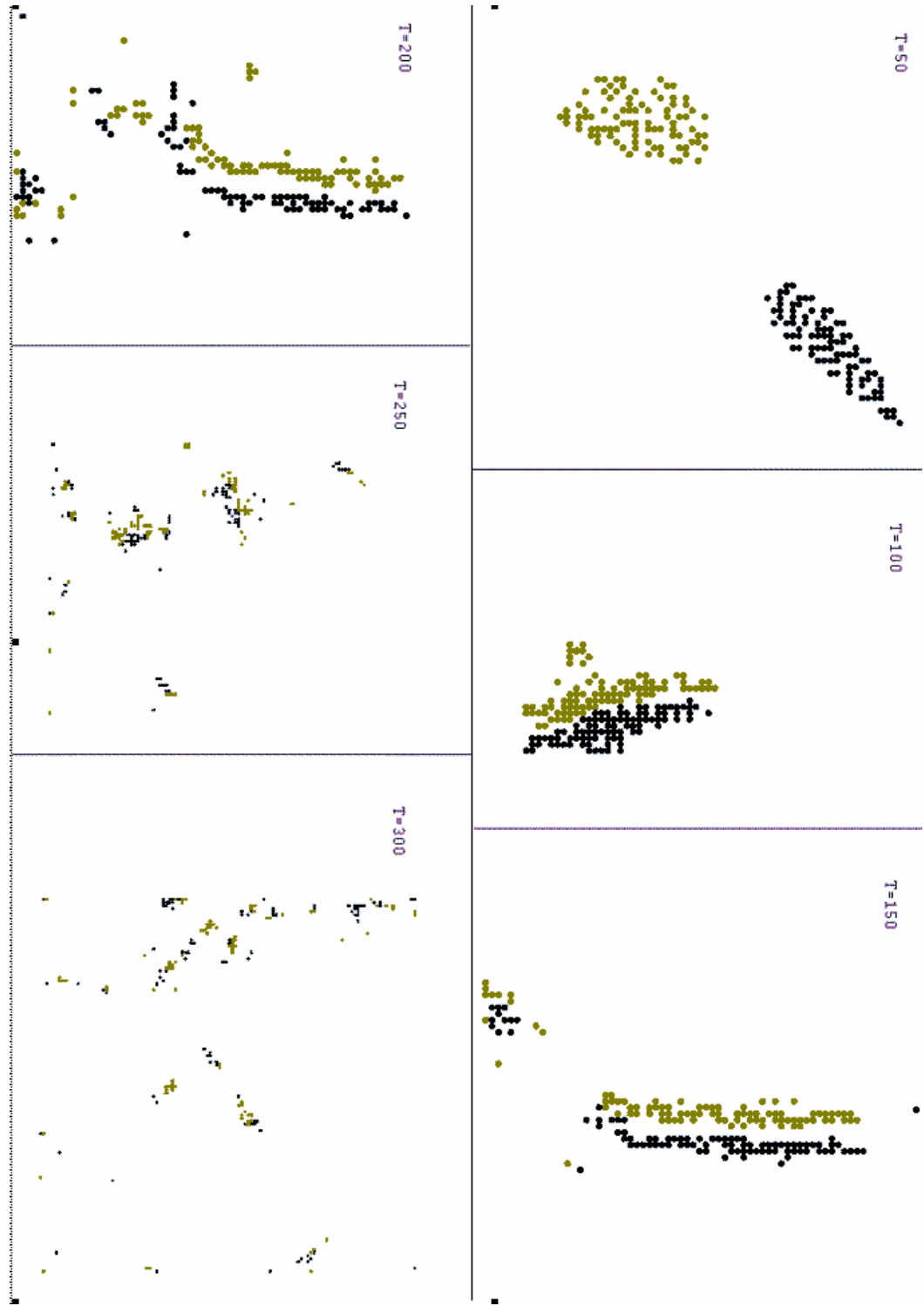

Figure 19: Evolution of the Einstein model demonstrating the "turbulent" transition point in an automaton battle, which in this case occurs at an ill-defined time between $T$ = 200 and $T$ = 250.

It is also apparent that the behaviour of the model is qualitatively similar for a broad range of values for these parameters. For fluid dynamics, velocity gradients and viscosity play important roles in determining turbulent behaviour, in particular, the onset of turbulence. This makes a nice analogy with the observed behaviour of ISAAC, in that the notional viscosity and force gradients determine the point where the behaviour of the model becomes "turbulent". One suspects that some kind of dimensionless quantity analogous to the Reynolds numbers in fluid dynamics exists also for ISAAC.

There are other analogies. For example, the onset of the turbulent state in fluid dynamics is both for practical purposes irreversible (without changes to the boundary conditions) and produces fractal spatial distributions of eddies. The transition between the laminar and turbulent phases cannot be predicted exactly, and occurs rapidly.

It must be noted that it is possible to pick parameters such that the simulation behaves as either one or the other of these "phases", so that there is no transition. We shall refer to the first phase as the "linear" phase, and the second as the "turbulent" phase.

For the linear phase of the battle shown in Figure 19, the outcome is likely to be quite similar to that given by conventional combat models, due to the fact that each side literally lines up and "shoots it out", at least to some point in time. Thus we expect the mean attrition rate during this time to be linearly dependent on the kill probability, $k$, so that, provided there are enough targets:

$$a = f(n,k) = \sum_n k_n = n\,k \qquad \text{if all } k_n \text{ are equal} \qquad (5.1)$$

where $a$ is the attrition rate, $n$ is the number of shooters, and $k_n$ is the kill probability for the $n$th shooter. The number of casualties is then:

$$C(t) = \int_0^t a(t)\,dt = f(t,k,n)$$

However, for the second phase of the battle, not all the red and blue automata are engaged in the combat at once. Furthermore, the distribution is continuously evolving in keeping with the two generic behaviours discussed above.

To describe the "turbulent" battlefield, let us further capitalise on the analogies with fluid dynamical turbulence. Note that when the Reynolds number is high for fluid turbulence, the mean square velocity spatial increment goes as $d^{2/3}$, as proposed Kolmogorov (1941) on the basis of dimensional requirements.

If a similar law applies to the ISAAC data obtained from the "turbulent" case discussed here, then it may be expected that:

$$\left\langle \left| n(T_0 + t) - n(T_0) \right|^2 \right\rangle \propto t^{2/3}$$

where we now consider temporal gradients, $n$ is the number of surviving automata, and the left-hand side is the "structure function" of the data $n$.

Clearly the rate of decline of automata numbers must be dependent on the kill probability. Since the structure function is dimensionless, we arrive at:

$$\left\langle \left| n(T_0 + t) - n(T_0) \right|^2 \right\rangle \propto f(n)\, k^{2/3}\, t^{2/3} \tag{5.2}$$

Thus for a time increment of arbitrary length, the square of the difference in automata numbers (and hence the attrition rate) depends on $k^{2/3}$, i.e. for "turbulent" combat the attrition rate is a non-linear function of kill probability.

This relationship was tested using 20 runs from the turbulent battlefield case for different values of $k$. Each run was stopped when the simulation reached the 50 per cent casualty level. The time taken was noted, and used as an attrition rate measure for that run. Figure 20 shows that the average attrition rate plausibly follows a $k^{1/3}$ theoretical curve. Note that the data points used were chosen so as to belong to the dominant population (as discussed in the preceding section), so that the "extreme" values were excluded.

This behaviour is significant. It appears that for the type of adaptive combat ISAAC produces, the value gained from increasing kill probability of the weapons (in terms of the size of the increase in attrition rate gained from an increase in kill probability) falls off rapidly towards to right-hand side of the curve, while at the other end, the attrition is much higher than might be expected given a linear dependence on kill probability. This suggests that an opponent with a relatively poor weapon performance may actually do much better than might be expected from a linear model.

That is done by increasing the degree to which they cluster together to increase their chances of destroying the enemy. It is remarkable that such a profound conclusion can be drawn from what appears to be a relatively simple model, without any explicit programming of tactics.

As an example of its significance, consider two forces with relatively low kill probability weapons (say Romans and barbarians). If the Romans have a significant kill probability advantage due to their superior weaponry, armour, and training, it is in their interest to try to force the barbarians to fight a "linear" battle by forming lines (a feasible approach with short-range weapons). Thus the Romans may expect a linear pay-off from their superior armaments. On the other hand, it is in the barbarians' interest to break the Roman lines and cause the battle to become "turbulent." Then, Figure 20 suggests that the Roman weaponry advantage would diminish markedly. (Of course, such an argument ignores the other advantages the Romans enjoyed from forming a "shield wall" using a linear formation. This was a standard tactic for civilised armies of antiquity.)

Another test of 5.2 is to examine the spectrum of the data. Using the Wiener-Khinchine relation the structure function can be related to the spectrum, so that it can be shown that $\left|\Delta t\right|^{\alpha-1}$ is a Fourier Transform pair with $\left|f\right|^{\alpha}$. Thus equation 5.2 implies that the spectrum of the data will be proportional to $f^{-5/3}$.

Figure 21 shows the spectrum of the data. The spectral slope obtained was $-1.67$, in good agreement with equation 5.2.

Though the agreement between the theoretical relation in equation 5.2 and the observed behaviour of the model appears good, some further justification for assuming the form of equation 5.2 is needed beyond simple analogies with fluid dynamics.

For a linear fight, the attrition rate is proportional to the product of the number of automata and the kill probability, as in equation 5.1. However, for the turbulent case, the automata do not simply "line up" and fight, rather, they split into various-sized clusters to fight. Consequently, suppose that for the turbulent case, the attrition rate is also a function of the probability of an automaton encountering a cluster in a given period of time. This might be done by finding the ratio of the number of clusters to the area of the battlefield.

However, determining what constitutes a cluster is not easy, since a single cluster can be viewed as a collect of smaller clusters, and so on. One way around this is to split the battlefield into arbitrarily sized boxes, and use the number of boxes containing automata as a measure of the number of clusters. But recall from the previous sections that the number of such boxes has a power-law dependence on the size of the box.

So, for the turbulent case, replace equation 5.1 with:

$$a = f(n,k) = f((l^{-2D}n_{cluster}),k) \qquad (5.3)$$

where we have replaced the total number of automata, $n$, with the number of boxes of side length $l$ (which is proportional to $l^{-2D}$) multiplied by the average number of automata per box, $n_{cluster}$, and $D$ is the box-counting fractal dimension of the distribution of forces across the battlefield.

Note that the size of the boxes can be related to time, since in a given time interval an automaton moving across the battlefield would on average map out a certain area, which depends on the speed it is moving. If we think of the automata playing the role of measuring sticks on the battlefield, then in a given time interval whether they will encounter automata clusters depends on the ratio of empty boxes to filled boxes, where the box is this case is the typical area each automaton maps out in that interval.

Thus the probability of an automaton encountering a box containing a cluster depends on the time interval, which is related to the length of the box by the proportionality constant $U$, where $U$ is the speed of the automaton.

Thus equation 5.3 may equivalently by written as:

$$a = f((t^{-2D}n_{cluster}),k)$$

for which dimensional requirements imply:

$$a \sim (t^{-2D}n_{cluster})\,k^{1-2D} \tag{5.4}$$

Since the attrition rate determines the difference between remaining automata numbers after a given temporal increment, similar arguments lead to a structure function of the form:

$$\left\langle \left| n(T_0+t)-n(T_0) \right|^2 \right\rangle \propto t^{2/3}\,n_{cluster}{}^2 k^{2/3} \tag{5.5}$$

where we have assumed $D = 1/3$. Note that the value of $D$ is obtainable by using the box-counting techniques discussed earlier. Equation 5.5 relates attrition rate to kill probability and number of participants (since $n_{total} \sim t^{2/3}\,n_{cluster}$) for the turbulent battlefield, just as the Lanchester equations do for the linear battlefield.

It is interesting to speculate whether this is the "new Lanchester equation" of the science of complexity.

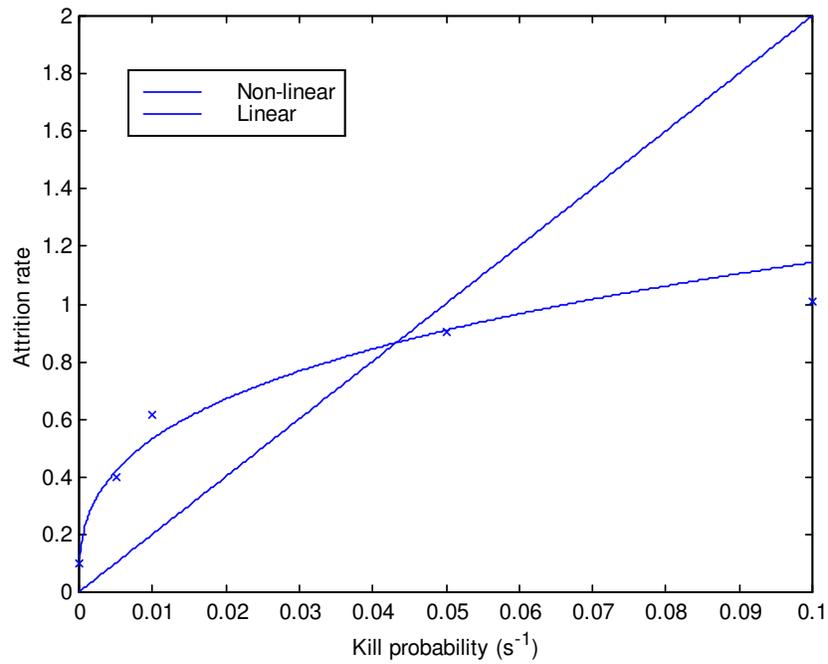

Figure 20: Plot of attrition rate versus kill probability for ISAAC data, compared with theoretical linear and non-linear curves.

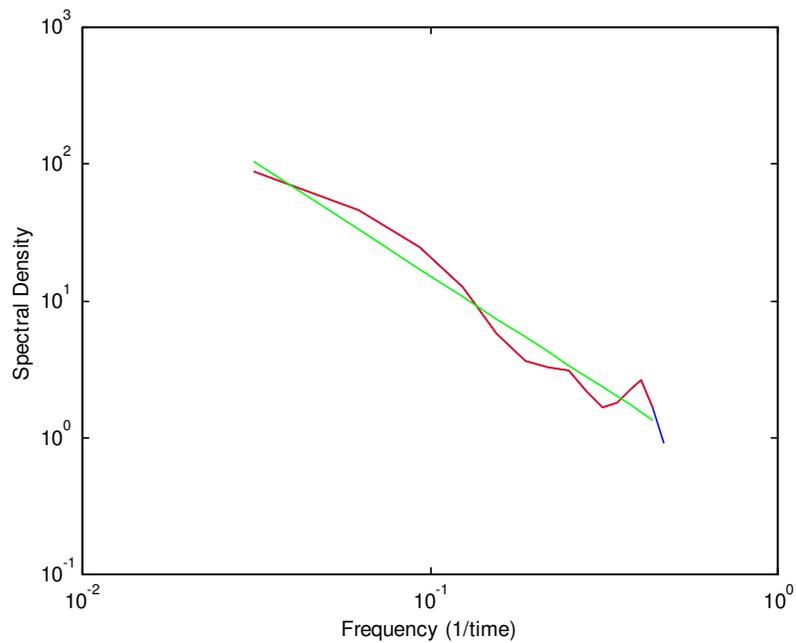

Figure 21: The spectrum of the function describing remaining red automata.

# 6. Distributions of combat outcomes

## 6.1 Napoleonic squares

In this section we consider what form the probability distributions for the outcomes of combat models should take. In particular, we are interested in the "scalability" of the results, in the sense of how increasing the size of the force involved affects the distribution of the outcomes. As was seen in the preceding sections, the attrition function for the ISAAC model showed temporal scaling (as evidenced by the existence of fractal dimensions and spectral power laws). In an earlier work (Lauren 1999), it was suggested that this temporal scaling may also lead the variance to scale in terms of unit size. In particular, it was suggested that the variance should demonstrate a power-law dependence on the size of the unit. In this way, the unit size is analogous to the "box size" used in equation 4.1, and the power-law slope related to a fractal dimension. This section investigates this, as well as the form for the probability distribution of combat outcomes for the ISAAC runs discussed in Section 2.

For conventional "Monte Carlo" combat models, the outcome of each run may be considered to be an independent random variable from an unknown probability distribution. Suppose we have a simple model with two equal forces on each side, representing, say, a simple small infantry battle. If the members of each side are able to cause a casualty to their opposition with a probability of $P_k$ with each time step, then for that time step the probability of causing a certain number of casualties will be described by a binomial distribution. Since the variance of the distribution changes as time progresses (since the variance of a binomial distribution is dependent on the number of remaining entities), after $x$ time steps the total casualties caused depends on potentially $x$ different intermediate binomial distributions.

Running such a model many times and recording the outcomes produces data with a finite mean and variance. Hence the distribution of the outcomes can be expected to be normal (Gaussian). This should certainly be so for the case shown in section 2.1. To test this, data was acquired by recording the remaining number of red automata after running the model many times for a set number of time steps. The number of time steps was chosen to be the mean time taken to reach 50 per cent casualties. The data collected for each of the cases was the result of thousands of runs on the Maui High Performance Computer Center. This data was used to produce a probability distribution for the outcomes.

Figure 22 compares the distributions for the ISAAC parameters used in section 2.1 with a Gaussian curve. The distributions are shown for the outcomes with 95, 50 and 10 automata on each side. All these probability distributions are similar to each other and either are Gaussian or very close to being so. It is noticeable that the standard deviation of the distribution increases as the size of the force decreases.

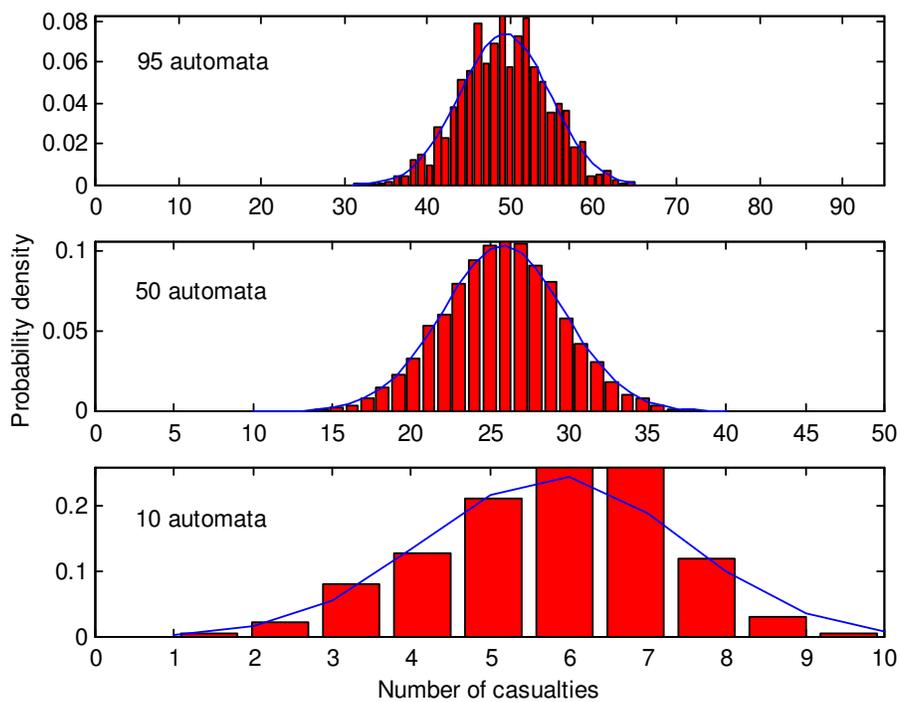

Figure 22: Distribution of outcomes for various-sized units using the Fluid_0D parameters.

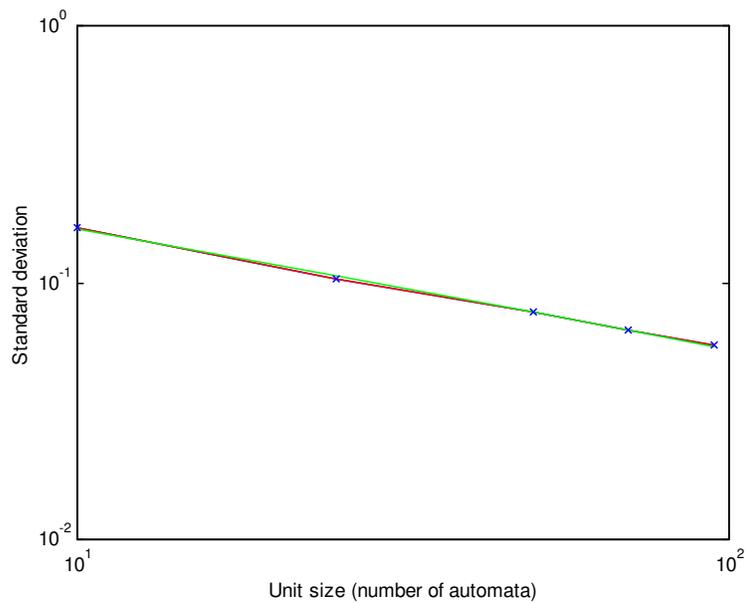

Figure 23: Scaling of the standard deviation with unit size for Fluid_0D parameters.

Figure 23 demonstrates that this unit-size scale dependence does indeed obey a power-law, as suggested in Lauren 1999. The slope here was $-0.47$.

## 6.2 Fluid battle

Before examining the probability distributions for the ISAAC runs discussed in section 2.2, the reader should first recall the qualitative observations made. In particular, that the behaviour of the simulation with 10 automata was markedly different from that with 95. That is, the former case displayed the "turbulent" dynamics discussed in the preceding section, while the latter case displayed "linear" dynamics.

Figure 24 shows the distributions of outcomes for the 10, 50, and 95 automata cases. Noticeably, the distributions for 50 and 95 automata are very similar to those in Figure 22.

The distribution for 10 is markedly different, and appears to be bimodal. This distribution is a product of a fundamentally different statistical nature of the dynamics for this number of automata, due to the difference between the linear and turbulent phases. In particular, the turbulent phase leads to data that contain extreme values that indicate the existence of multiple states and therefore multiple statistical populations. It was found that the shape of this distribution tended to be extremely sensitive to the number of time steps the model was run to. This was not the case for the distributions for 50 and 95 automata, which tended to be Gaussian provided the mean was not too close to either end of the range of possible survivor numbers.

Perhaps a more useful illustration of this difference in statistical character is shown in Figure 25. This compares the distribution of outcomes for 10 automata with the Fluid_1D parameter set, with the outcomes of the Fluid_0D parameters. Here the time limit for each run was chosen so that the mean casualty level was 15 per cent. Clearly, the ISAAC manoeuvre rules affect the distributions, giving the data a much longer tail.

Figure 26 shows the scaling of the standard deviation with unit size for the Fluid parameters. The power-law slope is $-0.48$. Noting that virtually all the data points for this plot were generated using parameters which lead to linear warfare, this suggests that the scaling law is not strongly dependent on the actual parameter value provided linear warfare occurs.

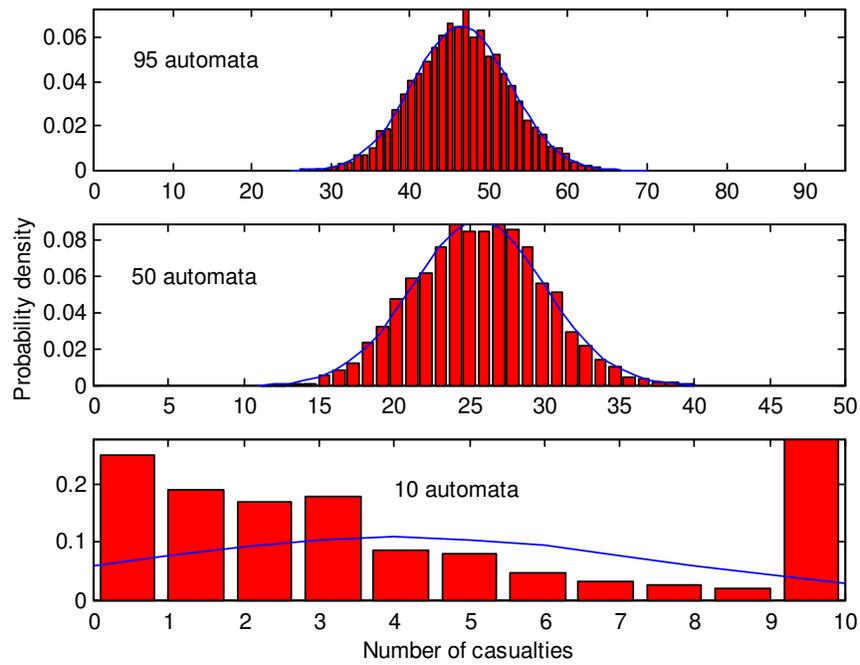

Figure 24: Distribution of outcomes for various-sized units using the Fluid_1D parameters.

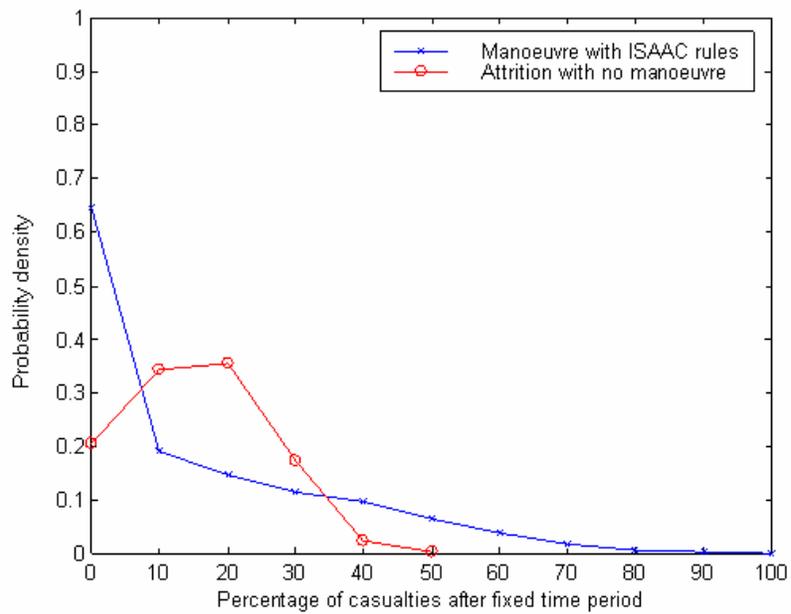

Figure 25: Distribution of battle outcomes for 10 automata with and without ISAAC manoeuvre rules (mean casualties is 15 per cent).

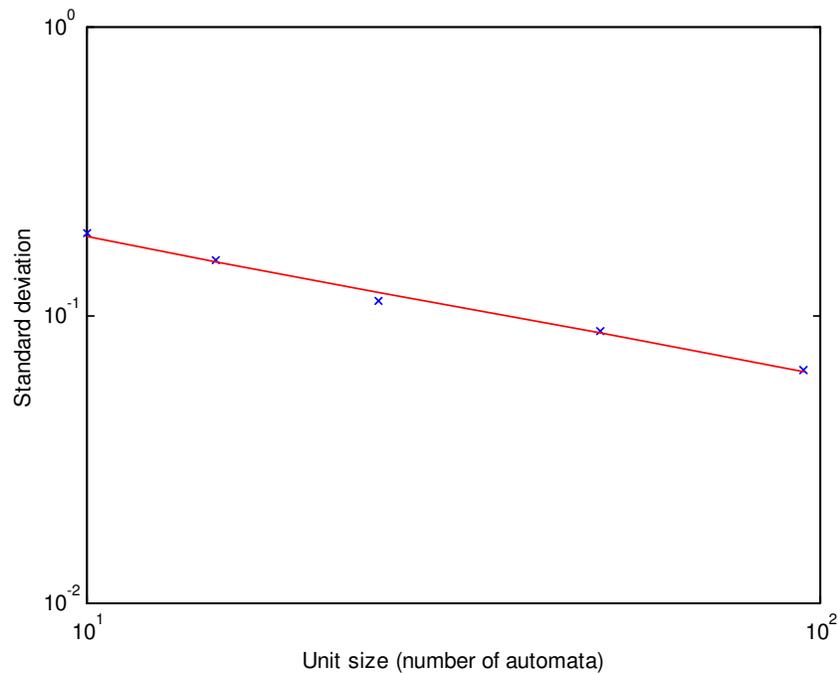

Figure 26: Scaling of the standard deviation with unit size for Fluid_1D parameters.

### 6.3 Effect of additional automata types

It was noted in section 2.3 that the adding of additional automata types to the model caused the dynamics to remain turbulent as the number of automata increased. (Note that the battle could also become turbulent for cases with larger numbers of a single automata type depending on the parameter set, as seen in section 5). This caused the distribution of times taken to reach 50 per cent casualties to contain extreme values. This leads us to suspect that the distribution of the outcomes should be non-Gaussian (since there are multiple states).

The distributions are shown in Figure 27. It is noticeable that the distribution becomes Gaussian for the case with 95 automata, but that it is not Gaussian for the 25 and 50 automata cases (in fact, appears somewhat bimodal, at least for the bottom case).

Figure 28 shows how the standard deviation scales with unit size. Here the slope was −0.82, significantly steeper than the −0.47 and −0.48 observed for the cases in sections 6.1 and 6.2. This suggests that the scaling of turbulent battle dynamics is markedly different from that of the linear case.

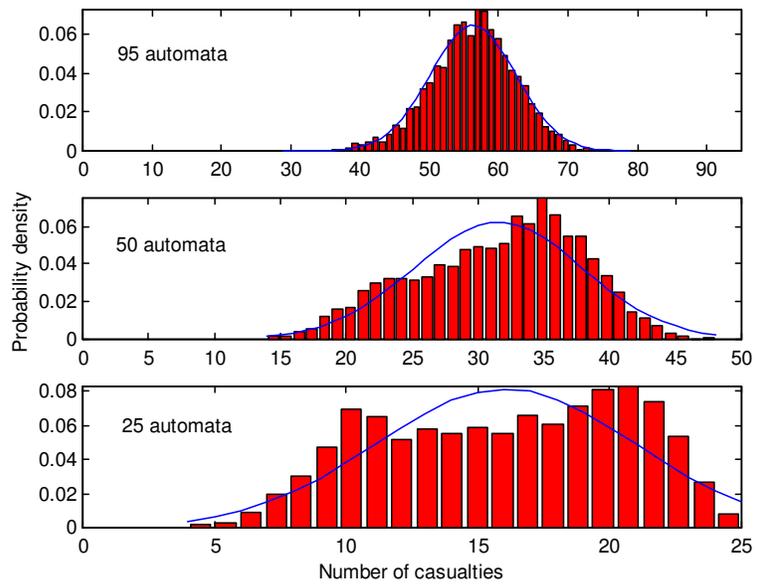

Figure 27: Distributions of outcomes for various-sized units using the Fluid_3D parameters.

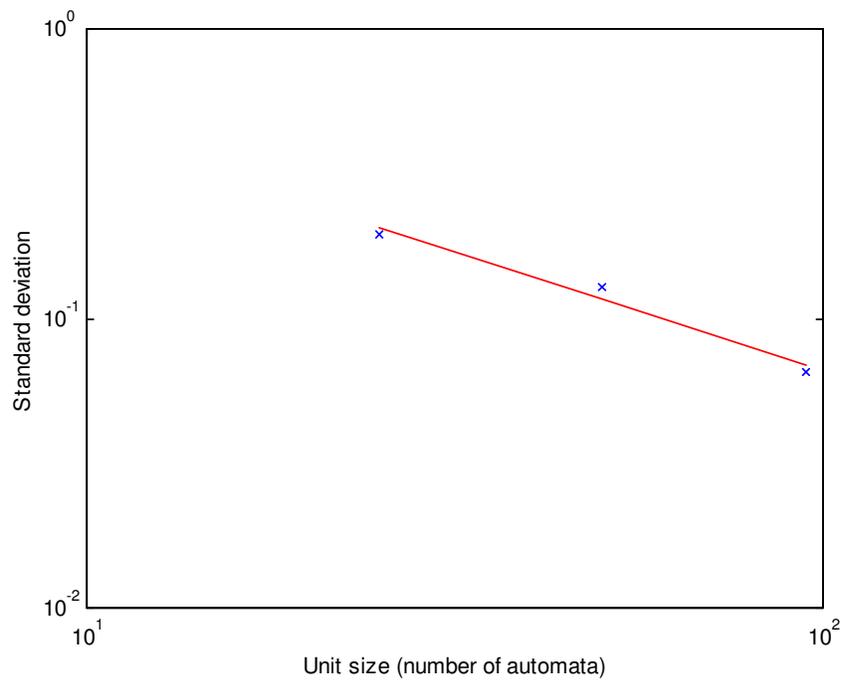

Figure 28: Scaling of the standard deviation as a function of unit size with the Fluid_3D parameters.

# 7 Conclusions

The aim of this paper was to demonstrate how manoeuvre affects the statistics of combat. The simplicity of a model like ISAAC allows us to explore such concepts without being tied to specific weapons systems. More significantly, the adaptive nature of the model allows a far greater range of possible outcomes for a given scenario to be explored than is currently feasible for conventional combat models.

It should also be noted that conventional combat models behave largely as a series of attrition-driven fights (attrition-driven in the sense that both sides usually fight to a certain level of attrition, then withdraw). By contrast, the ISAAC entities will only fight if conditions are suitable. This often means that the entities manoeuvre for extended periods to achieve this condition.

Given this behaviour, for two evenly balanced forces the dynamics of ISAAC appear to be drawn to two cases:

• Linear warfare: Where the forces literally fight in a line formation.

• Turbulent warfare: Where the formations are fluid and there exists a greater degree of freedom of movement for the entities.

Whether the battle evolves in one or the other of these two "phases" seems to depend largely on two things:

• The degree to which each force clusters together.

• The desire to minimise the local difference between friendly and enemy entities (i.e. minimise local force "gradients").

For the latter, there are two natural formations the entities may adopt during the battle which minimise these gradients:

• A line or curve, which for two even opponents spread out as thinly as possible; and

• A fractal distribution. Such spatial distributions are characteristic of fluid dynamical turbulence (hence the use of the term "turbulent" warfare). A fractal pattern reveals ever-more structure as it is examined on increasingly fine scale (i.e. large clusters are collections of smaller clusters, which are themselves collections of yet smaller clusters), so that in principle such a formation has an infinite surface area (though in practice doesn't here!), once again maximising the surface area between the two forces.

The statistics of linear warfare are quite similar to those of conventional combat models. This was seen by the fact that the statistical properties of both the attrition function and the distribution of outcomes were similar between this case and the case where the automata did not manoeuvre during the combat. However, it should be noted that the statistics of these two cases are not in general identical.

On the other hand, the turbulent warfare case displayed several traits not likely to be reproducible using conventional combat models. These were:

• A higher degree of correlation and "structure" in the attrition data.

• A greater degree of intermittency of casualties.

• The existence of extreme values, brought about by the model evolving into different end states. Here a state does not simply mean "the end result", but rather, a particular state is characterised by the model behaving in a certain way which is qualitatively very similar to some model runs, but distinctly different from others.

The first two of these traits imply that not only does attrition occur infrequently, but it occurs in clustered bursts.

Perhaps more interesting is the existence of extreme values in the statistics. These were evident when the model was used to either determine the time taken for one of the forces to reach a certain level of casualties, or when the model was run to a certain time step and the variance of the outcomes examined.

The possibility of identifying such extreme outcomes is intriguing because there is potential to understand what leads to these circumstances and try to exploit this information to make a substantial improvement in tactics or weapon/sensor mix. Additionally, it should be realised that what may seem to be a low-risk situation using conventional techniques may in fact contain extreme cases (with a much greater relative frequency than Gaussian statistics suggest) where the risk increases substantially. These cases must be identified and action taken to avoid such situations occurring in reality.

This raises an important point for the existing methods for analysing ISAAC data:

• Conventional statistical techniques (i.e. techniques which assume normal or Gaussian data) applied to ISAAC data will not characterise the extremities correctly;

and, therefore:

• The current data farming and corresponding visualisation tools that are used with ISAAC bury such extremities (see Hoffman and Horne 1998 for description of data farming).

The following procedure is suggested as an alternative to the such methods:

• Set up ISAAC to model a given concept.

• Run the model with these parameters many times, looking for extreme results.

• Examine the cases with the extreme results and try to determine what caused them.

• Identify the parameters associated with the cause of the extremities, then concentrate on these for further experimentation.

It should also be noted that the distribution of outcomes depended on the size of the force involved (in particular, the standard deviation). Notably, the statistics could also become non-Gaussian depending on the size of the force.

A particularly striking example was given in Figure 25. Here, the distribution of outcomes after a set time was shown for both the turbulent and static combat cases. The turbulent case had a markedly longer tail. It is interesting to compare this with research by the UK's Centre for Defence Analysis (Hall, A., Wright, W., and Young, M. 1997), which suggested that the distribution of outcomes of historical data obeys a lognormal curve (which, unlike Gaussian statistics, possesses a long "tail" of extreme values). Clearly, conventional combat models do not explain this historical distribution (since the results here suggest these should be Gaussian). However, the CDA report does not consider the possibility that the variance or shape of this distribution changes with unit size, or the nature of the battle.

Interestingly, the statistics of the structure function for the turbulent case appeared to be describable using an equation analogous to that for so-called inertial-range of turbulence, suggested by Kolmogorov. This not only predicted the slope of the power-law spectra (evidence of temporal correlation in the data), but suggested that the attrition rate was a non-linear function of weapon kill probability. This is extremely significant, because it suggests that the pay-off from improved weapon performance may be much smaller than a conventional model would suggest. This is directly related to the ability of each side to adapt to the situation it finds itself in.

Thus an important difference between ISAAC and conventional combat models is:

• Non-linearity of the pay-off of improved weapons performance.

That is to say, an $x$ per cent improvement in a weapon will not necessarily translate into the same percentage improvement in battle performance. In many cases, the improvement in battlefield performance will be significantly smaller.

The methods presented here are a way of determining the extent to which a model may exhibit complex behaviour. By obtaining the appropriate fractal parameters, it is in principle possible to replicate data indistinguishable from that produced using ISAAC, and which represents cases of differing levels of complexity. This may be especially useful for characterising the extent to which the data from each case exhibits fat probability tails. Additionally, understanding the nature of the "burstiness" of the data may be useful for force maintenance modelling, particularly if the time scales being considered are large enough to allow pools of replacements to be brought in.

Although the situations modelled here with ISAAC are not necessarily of a great deal of interest because of a lack of "realism", it serves to make the point that the statistics of ISAAC-like models are different from conventional models when the two are trying to model the same thing.

## Acknowledgements

The author wishes to acknowledge the useful comments and contributions of Dr G. M. Jacyna and Dr E. L. Blair of the MITRE Corporation, and Mr J. D. Sharpe, Mr S. A. Brown, Dr R. Marrett and Dr J. H. Buckingham of DOTSE.

**Appendix: Model parameters.**

<u>Fluid_0D</u>

| Weapons parameters: | Alive | Injured |
|---|---|---|
| Sensor range | 23 | 23 |
| Fire range | 23 | 23 |
| Movement | 0 | 0 |
| Defence | 1 | 1 |
| Max. simultaneous targets | 5 | 5 |
| Single-shot kill probability | 0.01 | 0.01 |

Note that since there is no movement, the values for personality parameters are not necessary.

Fluid_1D

| Weapons parameters: | Alive | Injured |
|---|---|---|
| Sensor range | 5 | 2 |
| Fire range | 2 | 2 |
| Movement | 2 | 2 |
| Defence | 1 | 1 |
| Max. simultaneous targets | 5 | 5 |
| Single-shot kill probability | 0.01 | 0.005 |
| Personality parameters: | | |
| Drawn to ... alive friends | 5 | 50 |
| alive foes | 40 | −60 |
| injured friends | 0 | 5 |
| injured foes | 60 | 0 |
| own flag | 0 | 0 |
| foe's flag | 5 | 5 |
| Meta-personality: | | |
| Minimum number required for advance towards foe | 5 | 10 |
| Maximum cluster size | 10 | 9 |
| Advantage required before engaging in combat | −7 | −5 |
| Range threshold for meta-personality parameters | 2 | 2 |

Note: Personality parameters are weightings by which an automaton is drawn to particular objective. Negative value indicates repulsion. Negative value for engaging in combat parameter indicates level by which an automaton can be outnumbered.

Fluid_3D

Infantry: As with Fluid_1D.

Armour:

| Weapons parameters: | Alive | Injured |
|---|---|---|
| Sensor range | 8 | 8 |
| Fire range | 7 | 7 |
| Movement | 3 | 3 |
| Defence | 1 | 1 |
| Max. simultaneous targets | 5 | 5 |
| Single-shot kill probability | 0.03 | 0.015 |
| Personality parameters: | | |
| Drawn to ... alive friends | 1 | 50 |
| alive foes | −40 | −60 |
| injured friends | 0 | 0 |
| injured foes | −20 | 0 |
| own flag | 0 | 0 |
| foe's flag | 5 | 0 |
| Meta-personality: | | |
| Minimum number required for advance towards foe | 1 | 1 |
| Maximum cluster size | 10 | 9 |
| Advantage required before engaging in combat | −7 | −5 |
| Range threshold for meta-personality parameters | 2 | 2 |

Artillery:

| Weapons parameters: | Alive | Injured |
|---|---|---|
| Sensor range | 20 | 20 |
| Fire range | 15 | 15 |
| Movement | 1 | 1 |
| Defence | 1 | 1 |
| Max. simultaneous targets | 5 | 5 |
| Single-shot kill probability | 0.1 | 0.01 |
| Personality parameters: | | |
| Drawn to ... alive friends | 1 | 50 |
| alive foes | −40 | −60 |
| injured friends | 0 | 0 |
| injured foes | −20 | 0 |
| own flag | 0 | 0 |
| foe's flag | 5 | 0 |
| Meta-personality: | | |
| Minimum number required for advance towards foe | 1 | 1 |
| Maximum cluster size | 10 | 9 |
| Advantage required before engaging in combat | −7 | −5 |
| Range threshold for meta-personality parameters | 2 | 2 |

Einstein_Fluid

| Weapons parameters: | Alive | Injured |
|---|---|---|
| Sensor range | 5 | 5 |
| Fire range | 3 | 3 |
| Movement | 2 | 2 |
| Defence | 1 | 1 |
| Max. simultaneous targets | 5 | 5 |
| Single-shot kill probability | 0.002 | 0.001 |
| Personality parameters: | | |
| Drawn to ... alive friends | 10 | 10 |
| alive foes | 40 | 40 |
| injured friends | 10 | 10 |
| injured foes | 40 | 40 |
| own flag | 0 | 0 |
| foe's flag | 25 | 25 |
| Meta-personality: | | |
| Minimum number required for advance towards foe | 3 | 4 |
| Maximum cluster size | 8 | 9 |
| Advantage required before engaging in combat | −2 | −4 |
| Range threshold for meta-personality parameters | 3 | 3 |

Transition:

| Weapons parameters: | Alive | Injured |
|---|---|---|
| Sensor range | 5 | 5 |
| Fire range | 1 | 1 |
| Movement | 1 | 2 |
| Defence | 1 | 1 |
| Max. simultaneous targets | 5 | 5 |
| Single-shot kill probability | 0.002 | 0.001 |
| Personality parameters: | | |
| Drawn to ... alive friends | 20 | 30 |
| alive foes | 40 | −40 |
| injured friends | 10 | 20 |
| injured foes | 40 | −10 |
| own flag | 0 | −10 |
| foe's flag | 0/5 | −10 |
| Meta-personality: | | |
| Minimum number required for advance towards foe | 10 | 4 |
| Maximum cluster size | 8 | 9 |
| Advantage required before engaging in combat | −2 | −4 |
| Range threshold for meta-personality parameters | 3 | 3 |

Note: Slash for the foe's flag personality reflects different values for red and blue forces.